\documentclass[12pt]{article}

\usepackage{epsf}
\usepackage[dvips]{graphicx}

\begin{document}

\begin{minipage}{14cm}
\vskip 0cm
\hspace*{10.2cm} {\bf JINR preprint}\\
\hspace*{10.5cm}{\bf E2-2006-34}\\
\hspace*{10.5cm}{\bf Dubna, 2006}\\
\end{minipage}
\vskip 2cm

\begin{center}
{\bfseries GENERALIZED Z-SCALING\\[2mm]
 IN PROTON-PROTON COLLISIONS AT HIGH ENERGIES}

\vskip 15mm
 I. Zborovsk\'{y}$^{a}$
 and  M. Tokarev$^{{b}}$

\vskip 0.2cm {\small
$^{(a)}${\it Nuclear Physics Institute,\\
Academy of Sciences of the Czech Republic, \\
\v {R}e\v {z}, Czech Republic} } \\
{\it E-mail: zborovsky@ujf.cas.cz}

\vskip 5mm {\small
$^{(b)}${\it Veksler and Baldin Laboratory of High Energies,\\
Joint Institute for Nuclear Research,\\
141980, Dubna, Moscow region, Russia}}\\
{\it E-mail: tokarev@sunhe.jinr.ru}
\end{center}

\vskip 5mm

\begin{center}
\begin{minipage}{150mm}
\centerline{\bf Abstract} New generalization of $z$-scaling in
inclusive particle production is proposed.  The scaling variable
$z$ is fractal measure which depends on kinematical
characteristics of the underlying sub-process expressed in terms
of the momentum fractions $x_1$ and $x_2$ of the incoming protons.
In the generalized approach, the $x_1$ and $x_2$ are functions of
the momentum fractions $y_a$ and $y_b$ of the scattered and recoil
constituents carried out by the inclusive particle and recoil
object, respectively. The scaling function $\psi(z)$ for charged
and identified hadrons produced in proton-proton collisions is
constructed. The fractal dimensions and heat capacity of the
produced medium entering definition of the $z$ are established to
obtain energy, angular and multiplicity independence of the
$\psi(z)$. The scheme allows unique description of data on
inclusive cross sections of charged particles, pions, kaons,
antiprotons and lambdas at high energies. The obtained results are
of interest to use $z$-scaling as a tool for searching for new
physics phenomena of particle production in high transverse
momentum and high multiplicity region at proton-proton colliders
RHIC and LHC.
\end{minipage}
\end{center}

\vskip 20mm
\begin{center}
  Presented at the Workshop of European Research Group\\
  on Ultra-Relativistic Heavy Ion Physics,\\
JINR, March 9-14, 2006, Dubna, Russia
\end{center}

\newpage

{\section{Introduction}}

Production of particles with high transverse momenta from
collision of hadrons and nuclei at sufficiently high energies has
relevance to constituent interactions at small scales. This regime
is of interest to search for new physical phenomena in elementary
processes such as quark compositeness \cite{Quarkcom}, extra
dimensions \cite{Extradim}, black holes \cite{Blackhole}, fractal
space-time \cite{Fracspace} etc. Other aspects of high energy
interactions are connected with small momenta of secondary
particles and high multiplicities. This regime has relevance to
collective phenomena of particle production. Search for new
physics in both regions is one of the main goals of investigations
at Relativistic Heavy Ion Collider (RHIC) at BNL and Large Hadron
Collider (LHC) at CERN. Experimental data  of particle production
can give constraints for different theoretical models. Processes
with high transverse momenta of produced particles are most
suitable for precise test of perturbative QCD. The soft regime is
of interest for verification of non-perturbative QCD and
investigation of phase transitions in non-Abelian theories.

Nucleus-nucleus interactions  are very complicated. In order to
understand their nature one often exploits phenomenology
and comparison with more simple proton-nucleus and proton-proton
collisions. Many approaches to description of particle production
are used to search for regularities reflecting general principles
in these systems at high energies \cite{Feynman}-\cite{Brodsky}.
One of the most basic principles is the self-similarity of hadron
production valid both in soft and hard physics. Other general
principles are locality and fractality which can be applied to
hard processes at small scales. The locality of hadronic
interactions follows from numerous experimental and theoretical
studies. These investigations have shown that interactions of
hadrons and nuclei can be described in terms of the interactions
of their constituents. Fractality in hard processes is specific
feature connected with sub-structure of the constituents. This
includes the self-similarity  over wide scale range.

Fractality of soft processes concerning the multi-particle
production was investigated by many authors. Fractality in
inclusive reactions with high-$p_T$ particles was considered for
the first time in the framework of $z$-scaling\cite{Z}. The
approach is based on principles of locality, self-similarity and
fractality. It takes into account fractal structure of the
colliding objects, interaction of their constituents and particle
formation. The scaling function $\psi(z)$ and the variable $z$ is
constructed via the experimentally measured inclusive cross
section $Ed^3\sigma/dp^3$ and the multiplicity density $dN/d\eta$.
In the original version, the construction was based on the
assumption that gross features of the inclusive particle
distribution for the inclusive reaction
\begin{equation}
M_{1}+M_{2} \rightarrow m_1 + X \label{eq:r1}
\end{equation}
at high energies can be described in terms of the corresponding
exclusive sub-process
\begin{equation}
(x_{1}M_{1}) + (x_{2}M_{2}) \rightarrow m_1 +
(x_{1}M_{1}+x_{2}M_{2} + m_2). \label{eq:r2}
\end{equation}
The $M_1$ and $M_2$ are masses of the colliding hadrons (or
nuclei) and $m_1$ is the mass of the inclusive particle. The
mass parameter $m_2$ is introduced in connection with internal conservation laws (for
isospin, baryon number, strangeness...). The $x_1$ and $x_2$ are
fractions of the incoming four-momenta $P_1$ and $P_2$ of the
colliding objects. The scaling variable $z$ was constructed as
fractal measure in terms of the fractions $x_1$ and $x_2$. It
depends on the nucleon anomalous dimension $\delta$ and the
average multiplicity density $dN/d\eta|_0$ of charged particles
produced in the central region of the interaction ($\eta=0$). The scaling for
non-biased collisions was established for single constant value of
$\delta$ as independence of the scaling function $\psi(z)$ on the
collision energy $\sqrt{s}$ and angle $\theta$ of the inclusive
particle.

The concept of $z$-scaling was generalized for various
multiplicities of produced particles \cite{ZZ}.  Relation of the
$z$-scaling to entropy $S$ and heat capacity $c$ of the colliding
system was established. The generalization was connected with
introduction of momentum fraction $y$ in the final state written
in the symbolic way
\begin{equation}
(x_{1}M_{1}) + (x_{2}M_{2}) \rightarrow (m_1/y) +
(x_{1}M_{1}+x_{2}M_{2} + m_2/y). \label{eq:r3}
\end{equation}
It was shown that generalized scaling represents regularity in
both soft and hard regime in proton-(anti)proton collisions over a
wide range of initial energies and multiplicities of the produced
particles. However, the generalization of the scaling for various
multiplicities was at the expense of the angular independence of
the scaling function observed for $y=1$.

In this paper we show that independence of the scaling function
$\psi(z)$ on the collision energy $\sqrt{s}$, multiplicity density
$dN/d\eta$ and angle $\theta$ can be restored simultaneously, if
two fractions $y_a$ and $y_b$ for the scattered constituent and
its recoil are introduced, respectively.

\vskip 0.5cm {\section{New generalization of $z$-scaling}}

We consider collision of extended objets (hadrons and nuclei) at
sufficiently high energies as an ensemble of individual
interactions of their constituents. The constituents are partons
in the parton model or quarks and gluons in the theory of QCD.
Single interaction of the constituents is illustrated in Fig.1.
Structures of the colliding objects are characterized by the
anomalous dimensions $\delta_1$ and $\delta_2$ in the space of
momentum fractions. Interacting constituents carry the fractions
$x_1$ and $x_2$ of the incoming objects. The sub-process is
considered as a binary collision with production of the scattered
and recoil constituents, respectively. The inclusive particle
carries the momentum fraction $y_a$ of the scattered constituent
which fragmentation is characterized by the anomalous dimension
$\epsilon_a$. Fragmentation of the recoil constituent is described
by the corresponding fractal dimension $\epsilon_b$ and momentum
fraction $y_b$.

Multiple interactions are considered to be similar. The property
is manifestation of the self-similarity of hadronic interaction at
the constituent level. The interactions are governed by local
energy-momentum conservation law. The self-similarity at small
scales reflects fractality of the interaction objects and their
constituents characterized by the corresponding fractal
dimensions. The fractality concerns parton content of the
composite structures involved.

\vskip 0.5cm {\subsection{Locality, self-similarity and
fractality}}

The idea of $z$-scaling is based on the assumption
\cite{Stavinsky} that gross features of inclusive particle
distribution of the reaction (\ref{eq:r1}) can be described at
high energies in terms of the kinematic characteristics of the
corresponding constituent sub-processes. We consider the
sub-process as binary collision
\begin{equation}
(x_{1}M_{1}) + (x_{2}M_{2}) \rightarrow (m_1/y_a) +
(x_{1}M_{1}+x_{2}M_{2} + m_2/y_b) \label{eq:r4}
\end{equation}
of the constituents $(x_{1}M_{1})$ and $(x_{2}M_{2})$ resulting in
the scattered $(m_1/y_a)$ and recoil $(x_{1}M_{1}+x_{2}M_{2} +
m_2/y_b)$ constituents in the final state. The inclusive particle
with the mass $m_1$ and the 4-momentum $p$ carries out the
fraction $y_a$ of the 4-momentum of the scattered constituent. Its
counterpart ($m_2$), moving in the away side direction, carries
out the 4-momentum fraction $y_b$ of the produced recoil. The
binary sub-process satisfies the energy-momentum conservation
written in the form
\begin{equation}
(x_1P_1+x_2P_2-p/y_a)^2 =(x_1M_1+x_2M_2+m_2/y_b)^2.
\label{eq:r5}
\end{equation}
The equation is expression of the locality of hadron interaction
at constituent level. It represents kinematical constraint on the
fractions $x_1$, $x_2$, $y_a$ and $y_b$.

The self-similarity of hadron interactions reflects property
that hadron constituents and their
interactions are similar. This is connected with dropping of
certain dimensional quantities out of description of physical
phenomena. The self-similar solutions are constructed in terms of
self-similarity parameters. We search for the solution
\begin{equation}
\psi(z) ={1\over{N\sigma_{in}}}{d\sigma\over{dz}}
\label{eq:r6}
\end{equation}
depending on single self-similarity variable $z$. Here
$\sigma_{in}$ is the inelastic cross section of the reaction
(\ref{eq:r1}) and $N$ is the average particle multiplicity. The
variable $z$ is specific dimensionless combination of quantities
which characterize particle production in high energy inclusive
reactions. It depends on  momenta and masses of  the colliding and
inclusive particles, structural parameters of the interacting
objects and dynamical characteristics of the produced system. We
define the self-similarity variable $z$ in the form
\begin{equation}
z =z_0 \Omega^{-1}
\label{eq:r7}
\end{equation}
where
\begin{equation}
\Omega(x_1,x_2,y_a,y_b)=
(1-x_1)^{\delta_1}(1-x_2)^{\delta_2}(1-y_a)^{\epsilon_a}(1-y_b)^{\epsilon_b}.
\label{eq:r8}
\end{equation}

The  variable $z$ has character of a fractal measure. For a given
reaction (\ref{eq:r1}), its finite part $z_0$ is proportional to
the transverse kinetic energy of the constituent sub-process
consumed on the production of the inclusive particle ($m_1$) and
its counterpart ($m_2$). The divergent factor $\Omega^{-1}$
describes resolution at which the sub-process can be singled out
of this reaction. The $\Omega(x_1,x_2,y_a,y_b)$ is relative number
of parton configurations containing the incoming constituents
which carry the fractions $x_1$ and  $x_2$ of the momenta $P_1$
and $P_2$ and the outgoing constituents which fractions $y_a$ and
$y_b$ are carried out by the inclusive particle $(m_1)$ and its
counterpart $(m_2)$, respectively. The $\delta_1$ ($\delta_2$) and
$\epsilon_a$ ($\epsilon_b$) are anomalous fractal dimensions of
the incoming and fragmenting objects, respectively. For inelastic
proton-proton collisions we have $\delta_1=\delta_2\equiv\delta$.
We also assume that fragmentation of the scattered and recoil
constituents is governed by the same anomalous dimensions
$\epsilon_a=\epsilon_b\equiv\epsilon$.

Common property of fractal measures is their divergence with the
increasing resolution
\begin{equation}
z(\Omega) \rightarrow \infty \ \ \ \ \ \ \ \ if  \ \ \ \ \
\Omega^{-1} \rightarrow \infty.
\label{eq:r9}
\end{equation}
For the infinite resolution, all momentum fractions become unity
($x_1=x_2=y_a=y_b=1$) and $\Omega=0$. The kinematical limit
corresponds to the fractal limit $z=\infty$.

\vskip 0.5cm {\subsection{Principle of minimal resolution}}

The momentum fractions $x_1$, $x_2$, $y_a$ and $y_b$ are
determined form principle of minimal resolution of the fractal
measure $z$. The principle states that resolution $\Omega^{-1}$
should be minimal with respect to all binary sub-processes
(\ref{eq:r4}) in which the inclusive particle $m_1$ with the
momentum $p$ can be produced. This singles out the underlying
interaction of the constituents. The momentum fractions $x_{1}$,
$x_{2}$, $y_a$ and $y_b$ are found from minimization of
$\Omega^{-1}(x_{1},x_{2},y_a,y_b)$,
\begin{equation}
\frac{\partial\Omega(x_i,y_j)}{\partial x_1} = 0, \ \ \ \ \ \
\frac{\partial\Omega(x_i,y_j)}{\partial x_2} = 0, \ \ \ \ \ \
\frac{\partial\Omega(x_i,y_j)}{\partial y_a} = 0, \ \ \ \ \ \
\frac{\partial\Omega(x_i,y_j)}{\partial y_b} = 0,
\label{eq:r10}
\end{equation}
taking into account the
energy-momentum conservation (\ref{eq:r5}).
The momentum fractions $x_1$ and $x_2$ can be decomposed as follows
\begin{equation}
x_1=\lambda_1+\chi_1(\alpha), \ \ \ \ \ \ \
x_2=\lambda_2+\chi_2(\alpha).
\label{eq:r11}
\end{equation}
The parameter $\alpha=\delta_2/\delta_1$ is ratio of the
anomalous fractal dimensions of the colliding objects.
Using the decomposition, the expression (\ref{eq:r4}) can be
rewritten to the symbolic form
\begin{equation}
x_1+x_2\rightarrow (\lambda_1+\lambda_2) + (\chi_1+\chi_2).
\label{eq:r12}
\end{equation}
This relation means that $\lambda$-parts of the interacting
constituents contribute to the production of the inclusive
particle, while the $\chi$-parts are responsible for the creation
of its recoil. The $\lambda's$ are functions of $y_a$ and $y_b$
\begin{equation}
\lambda_1=\kappa_1/y_a+\nu_1/y_b, \ \ \ \ \ \
\lambda_2=\kappa_2/y_a+\nu_2/y_b, \ \ \ \ \ \
\lambda_0=\bar{\nu}_0/y_b^2-\nu_0/y_a^2,
\label{eq:r15}
\end{equation}
where
\begin{equation}
\kappa_1=\frac{(P_2p)}{(P_1P_2)-M_1M_2}, \ \ \ \
\kappa_2=\frac{(P_1p)}{(P_1P_2)-M_1M_2}, \ \ \ \
\label{eq:r16}
\end{equation}
\begin{equation}
\nu_1=\frac{M_2m_2}{(P_1P_2)-M_1M_2}, \ \ \ \
\nu_2=\frac{M_1m_2}{(P_1P_2)-M_1M_2}, \ \ \ \
\label{eq:r17}
\end{equation}
\begin{equation}
\nu_0=\frac{0.5m_1^2}{(P_1P_2)-M_1M_2}, \ \ \ \
\bar{\nu}_0=\frac{0.5m_2^2}{(P_1P_2)-M_1M_2}. \ \ \ \
\label{eq:r18}
\end{equation}
The $\chi's$ are expressed via $\lambda's$ as follows
\begin{equation}
\chi_1=\sqrt{\mu_1^2+\omega_1^2}-\omega_1, \ \ \ \
\chi_2=\sqrt{\mu_2^2+\omega_2^2}+\omega_2,
\label{eq:r13}
\end{equation}
where
\begin{equation}
\mu_1^2=(\lambda_1\lambda_2+\lambda_0)\alpha\frac{1-\lambda_1}{1-\lambda_2},
\ \ \ \
\mu_2^2=(\lambda_1\lambda_2+\lambda_0)\alpha^{-1}\frac{1-\lambda_2}{1-\lambda_1},
\label{eq:r14}
\end{equation}
and $\omega_i=\mu_iU$ $(i=1,2)$.

The quantity
\begin{equation}
U = \frac{\alpha-1}{2\sqrt{\alpha}}\xi
\label{eq:r19}
\end{equation}
has physical meaning of longitudinal component of the "structural
4-velocity" \cite{SRT}. It is function of $\alpha$ and the
kinematical factor
\begin{equation}
\xi=\sqrt{\frac{\lambda_1\lambda_2+\lambda_0}{(1-\lambda_1)(1-\lambda_2)}},
\label{eq:r20}
\end{equation}
$(0\le\xi\le 1)$. The $\xi$ characterizes kinematical scale of the underlying
constituent interaction.

Solution of the system (\ref{eq:r10}) with the
condition (\ref{eq:r5}) can be obtained by searching for the
unbounded maximum of the function
\begin{equation}
F(y_a,y_b)\equiv\Omega\left(x_1(y_a,y_b),x_2(y_a,y_b),y_a,y_b\right)
\label{eq:r21}
\end{equation}
of two independent variables $y_a$ and $y_b$. Here $x_i(y_a,y_b)$ are given
explicitly by the expressions (\ref{eq:r11}). There exists single
maximum of the function $F(y_a,y_b)$ in the allowed kinematical
region and we determined it numerically.

\vskip 0.5cm
{\subsection{Scaling variable $z$}}

Search for an adequate, physically meaningful but still sufficiently simple form of the
self-similarity parameter $z$ plays a crucial role in our approach.
We define the scaling variable $z$ in the form
\begin{equation}
z = \frac{s^{1/2}_{\bot}}{(dN/d\eta|_0)^c \cdot
m}\cdot\Omega^{-1}.
\label{eq:r22}
\end{equation}
Here $m$ is a mass constant which we fix at the nucleon mass. The
transverse kinetic energy of the constituent sub-process consumed
on the production of the inclusive particle ($m_1$) and its
counterpart ($m_2$) is determined by the formula
\begin{equation}
s^{1/2}_{\bot}= T_a+T_b,
\label{eq:r23}
\end{equation}
where
\begin{equation}
T_a=y_a(s^{1/2}_{\lambda}-M_1\lambda_1-M_2\lambda_2)-m_1,
\label{eq:r24}
\end{equation}
\begin{equation}
 T_b= y_b(s^{1/2}_{\chi} -M_1\chi_1-M_2\chi_2)- m_2
\label{eq:r25}.
\end{equation}
For more details see the Appendix. The terms
\begin{equation}
s^{1/2}_{\lambda}=\sqrt{(\lambda_1P_1+\lambda_2P_2)^2},     \ \ \ \ \
s^{1/2}_{\chi}=\sqrt{(\chi_1P_1+\chi_2P_2)^2}
\label{eq:r26}
\end{equation}
represent energy for production of the scattered constituent and
its recoil, respectively. The boundaries of the range of the
variable $z$ are 0 and $\infty$. They are accessible at any
collision energy.

The $dN/d\eta|_0$ is average multiplicity density of charged particles produced
in the central region of the reaction (\ref{eq:r1}) at
pseudorapidity $\eta=0$. It depends on state of the produced
medium in the colliding system. The parameter $c$ characterizes
properties of this medium. The quantity
\begin{equation}
{\it W}=(dN/d\eta|_0)^c\cdot\Omega
\label{eq:r27}
\end{equation}
is proportional to all parton and hadron configurations of the colliding system
which can contribute to production of the inclusive particle with the momentum $p$.
The scaling variable (\ref{eq:r22}) is proportional to the ratio
\begin{equation}
z \sim \frac{s^{1/2}_{\bot}}{{\it W }}
\label{eq:r28}
\end{equation}
of the transverse kinetic energy $s^{1/2}_{\bot}$  and maximal
number of the configurations $W$.

\vskip 0.5cm
{\subsection{Scaling variable $z$ and entropy ${\it S}$}}

According to statistical physics, entropy of a system is given by
number of all statistical states $W$ of the system as follows
\begin{equation}
{\it S} = \ln {\it W}.
\label{eq:r29}
\end{equation}
In thermodynamics, entropy for ideal gas is determined by the formula
\begin{equation}
{\it S} = c_V\ln {T}+ R\ln {V} + const.
\label{eq:r30}
\end{equation}
The $c_V$ is heat capacity and $R$ is universal constant. The temperature $T$
and the volume $V$ characterize state of the system. Using
(\ref{eq:r27}) and (\ref{eq:r29}), we can write
\begin{equation}
{\it S} = c\ln {\left[dN/d\eta|_0\right]}+
\ln{[(1\!-\!x_1)^{\delta_1}(1\!-\!x_2)^{\delta_2}
(1\!-\!y_a)^{\epsilon_a}(1\!-\!y_b)^{\epsilon_b}]}.
\label{eq:r31}
\end{equation}
Exploiting analogy between Eqs. (\ref{eq:r30}) and (\ref{eq:r31}), we interpret
the parameter $c$ as "heat capacity" of the produced medium.
The multiplicity density $dN/d\eta|_0$
 has physical meaning of "temperature" of the colliding system.
The second term in Eq. (\ref{eq:r31}) depends on volume in
space of the momentum fractions $\{x_1,x_2,y_a,y_b\}$.
This analogy emphasizes once more interpretation of the parameters
$\delta_1$, $\delta_2$, $\epsilon_a$ and $\epsilon_b$ as fractal dimensions.
As seen from Eq. (\ref{eq:r31}), entropy of the colliding system increases
with the multiplicity density and decreases with increasing
resolution $\Omega^{-1}$.

Let us note that entropy (\ref{eq:r29}) of a system is determined up to an
arbitrary constant $\ln W_0$. Dimensional units
entering definition of the entropy can be included within this
constant. Therefore, it allows us to make analogy between
dimensionless $dN/d\eta|_0$ and the temperature. This degree of freedom is
connected with the transformation
\begin{equation}
z\rightarrow  W_0\cdot z, \ \ \ \
\psi \rightarrow W_0^{-1}\cdot \psi.
\label{eq:r32}
\end{equation}
In such a way the scaling variable and the scaling function
are determined up to an arbitrary multiplicative constant.

\vskip 0.5cm
{\subsection{Scaling function $\psi(z)$}}

The scaling function $\psi(z)$  is expressed in terms of the experimentally
measured inclusive invariant cross section $Ed^3\sigma/dp^3$, multiplicity density
$dN/d\eta$ and the total inelastic cross section $\sigma_{in}$.
Exploiting the definition (\ref{eq:r6}) one can obtain the expression
\begin{equation}
\psi(z) = -{ { \pi s} \over { (dN/d\eta) \sigma_{in}} } J^{-1} E {
{d^3\sigma} \over {dp^3}  }.
\label{eq:r33}
\end{equation}
Here $s$ is the centre-of-mass collision energy squared and
\begin{equation}
J = \frac{\partial \eta }{\partial\kappa_1}\frac{\partial z
}{\partial\kappa_2}- \frac{\partial \eta
}{\partial\kappa_2}\frac{\partial z }{\partial\kappa_1}
\label{eq:r34}
\end{equation}
is the corresponding Jacobian.
Angular properties the $\psi(z)$ showed that the scaling
is valid rather in pseudorapidity than in rapidity (see Appendix).
The function $\psi(z)$ is
normalized as follows
\begin{equation}
\int_{0}^{\infty} \psi(z) dz = 1.
\label{eq:r35}
\end{equation}
The relation allows us to interpret the $\psi(z)$ as a probability
density to produce inclusive particle with the corresponding value
of the variable $z$.

\vskip 0.5cm {\section{Properties of the scaling function $\psi(z)$}}

Let us investigate properties of $z$ presentation of experimental data obtained in
proton-proton collisions at high energies.

\vskip 0.5cm {\subsection{Energy independence of $\psi(z)$}}

We analyze experimental data \cite{FNAL1}-\cite{Adams} on
inclusive hadron ($h^{\pm}, \pi^-, K^-$ and $\bar{p}$) production
in minimum-biased proton-proton collisions. The data on inclusive
cross sections were measured in central rapidity region at FNAL,
ISR and RHIC energies $\sqrt{s}=19-200$~GeV.

The energy dependence of the charged hadron spectra on the
transverse momentum is shown in Fig. 2(a). The  distributions
cover the range up to $p_T\simeq 10$~GeV/c. The cross sections
change within the range of 12 orders of magnitudes. Strong
dependence of the spectra on the collision energy $\sqrt{s}$
increases with transverse momentum. Figure 2(b) shows
$z$-presentation of the same data. The scaling variable $z$
depends on the average multiplicity density of charged particles
produced in the central pseudorapidity region of the collision. We
have used experimentally measured values of $dN/d\eta|_{\eta=0}$
\cite{Ward} for minimum-biased collisions  in the analysis of
energy and angular properties of $\psi(z)$. Independence of the
scaling function $\psi(z)$ on collision energy $\sqrt{s}$ is found
for the constant values of the parameters $c=0.25, \delta=0.5$ and
$\epsilon=0.2$. The form of $\psi(z)$ manifests two regimes of
particle production. The hard regime is characterized by the power
law $\psi(z)\sim z^{-\beta}$ at large $z$. Soft processes
correspond to  behavior of the $\psi(z)$ at small $z$. Slope of
the scaling curve decreases with $z$ in this region.

Invariant cross sections for $\pi^-$-meson production as function
of the collision energy and transverse momentum are plotted in
Fig. 3(a). The spectra were measured over a wide transverse
momentum range $p_{T}=0.1-10$~GeV/c. The cross sections change
from $10^2$ to $10^{-10}$~mb/GeV$^2$. Strong dependence of
the pion spectra on $\sqrt{s}$ as for charged hadrons is observed. The
$z$-presentation of the same data is shown in Fig. 3(b).
For pions (as well as for all other types of the particles - kaons,
antiprotons, ... ), the dependence of $z$  on the charged particle
multiplicity density $dN/d\eta|_{\eta=0}$ have been used in the
formula (\ref{eq:r22}). Let us stress that, unlike this, the
scaling function (\ref{eq:r33}) is normalized to the multiplicity
density of pions. Independence of the scaling function for pions on $\sqrt{s}$
was obtained at $c=0.25, \delta=0.5$ and $\epsilon=0.2$ as for charged hadrons.
Shape of the $\psi(z)$ is similar in both cases, as well.

Transverse momentum spectra for $K^-$-meson production are shown in
Fig. 4(a). The cross sections were measured in the range
$p_{T}=0.1-8$~GeV/c. Data for $K^0_s$-mesons obtained by the STAR Collaboration at RHIC
are presented in the Fig. 4(a) as well.
The $K$-meson spectra demonstrate strong dependence on
the collision energy $\sqrt{s}$.
The corresponding scaling function $\psi(z)$ is depicted in Fig. 4(b).
Independence of the $\psi(z)$ on $\sqrt{s}$
is restored at $c=0.25, \delta=0.5$ and $\epsilon=0.3$.
Similar features of $p_T$ and $z$ presentations of experimental data on antiproton
production are presented in Figs. 5(a) and 5(b). The energy independence of
$\psi(z)$ for antiprotons is established at $c=0.25, \delta=0.5$ and $\epsilon=0.35$.

As a result we can conclude that energy independence of the
scaling function $\psi(z)$ is valid for different types of hadrons
in a wide range of centre-of-mass energy $\sqrt{s}$ and transverse
momentum $p_T$.

\vskip 0.5cm {\subsection{Angular independence of $\psi(z)$}}

We analyze experimental data \cite{ISR,CHLM} on angular dependence
of negative hadrons (pions, kaons and antiprotons) measured at ISR
energies. The data were measured both in the central and
fragmentation regions. Results of our analysis is demonstrated at
the energy $\sqrt{s}=53$~GeV.

Invariant cross sections for $\pi^-$-meson production as function
of the centre-of-mass angle $\theta$ and transverse momentum are
shown in Fig. 6(a). The angles cover the range $\theta=3^0-90^0$.
The central and fragmentation regions are distinguished by
different behavior of differential cross sections. The
$z$-presentation of the same data is demonstrated in Fig. 6(b).
The charged hadron multiplicity density  $dN/d\eta|_{\eta=0}$
represents angular independent factor in the definition of the
variable $z$. Contrary to this, the scaling function
(\ref{eq:r33}) is normalized to the multiplicity density
$dN/d\eta$ of pions depending on the angle $\theta$. Angular and
energy independence of the scaling function for pions was obtained
at the same values of $c=0.25, \delta=0.5$ and $\epsilon=0.2$. The
function $\psi(z)$ for small $\theta$ is sensitive to the value of
$m_2$. This parameter is determined from the corresponding
exclusive reaction
\begin{equation}
p+p\rightarrow \pi^- +p+\Delta^{++}.
\label{eq:r36}
\end{equation}
The reaction is limiting case of the sub-process (\ref{eq:r4}) for
$x_1=x_2=y_a=y_b=1$. According to Eqs. (\ref{eq:r5}) and
(\ref{eq:r36}), we obtain $m_2=m(\Delta^{++})-m(p)=0.3$~GeV. This
value was used in our analysis for inclusive $\pi^-$-meson
production.

Transverse momentum spectra for $K$-mesons and antiprotons produced in
$pp$ collisions at different angles are shown Fig. 7(a) and 8(a).
In addition to ISR data at $\sqrt{s}=53$~GeV, the data from RHIC at $\sqrt{s}=200$~GeV
are shown as well. The angular dependence of the spectra demonstrate
strong difference between central and fragmentation regions.
The corresponding function $\psi(z)$  for kaons and antiprotons
is plotted in Figs. 7(b) and 8(b), respectively.
For both particles, the charged hadron multiplicity density  $dN/d\eta|_{\eta=0}$
represents angular independent factor in the definition of the variable $z$.
The scaling function (\ref{eq:r33}) is normalized to the angular dependent multiplicity
density $dN/d\eta$ of kaons and antiprotons, respectively.

Independence of the $\psi(z)$ on the angle $\theta$
is obtained at $c=0.25, \delta=0.5$ and $\epsilon=0.3$ for kaons and
$c=0.25, \delta=0.5$ and $\epsilon=0.35$ for antiprotons.
Values of these parameters allow us to obtain simultaneously angular and energy
independence of the scaling function.
Like in the case of $\pi^-$-mesons, the function $\psi(z)$ for kaons and antiprotons
is sensitive to the value of $m_2$ at small angles $\theta$.
The corresponding exclusive reactions
\begin{equation}
p+p\rightarrow K^- +p+p+K^+,
\label{eq:r37}
\end{equation}
\begin{equation}
p+p\rightarrow \bar{p}+p+p+p
\label{eq:r38}
\end{equation}
were used to determine the parameter $m_2$. Exploiting Eq.
(\ref{eq:r5}) for $x_1=x_2=y_a=y_b=1$ and Eqs. (\ref{eq:r37}) and
(\ref{eq:r38}), we obtain $m_2=m(K^{+})=0.5$~GeV and
$m_2=m(p)=0.94$~GeV for the inclusive production of $K^-$-mesons
and antiprotons, respectively. These values were used in our
analysis.

\vskip 0.5cm {\subsection{Multiplicity independence of $\psi(z)$ }}

The STAR Collaboration obtained the new data \cite{Gans} on multiplicity dependence of
the inclusive spectra of charged hadrons produced
in $pp$ collisions in the central rapidity range $|\eta|<0.5$ at the energy $\sqrt s
= 200$~GeV.  The transverse momentum distributions were measured
up to 9.5~GeV/c using different multiplicity selection criteria.
Figure 9(a) demonstrates strong dependence of the spectra on
multiplicity density at $dN/d\eta=2.5, 6.0$ and 8.0.
The same data are presented in Figure 9(b) in the scaling form.
The scaling function $\psi(z)$ changes over 6 orders of magnitude
in the range of $z=0.2-10$. The independence of $\psi(z)$ on
multiplicity density $dN/d\eta$ is obtained. The result gives
strong restriction on the parameter $c$. It was found to be
$c=0.25$.

The STAR Collaboration measured the multiplicity dependence of the
$K^0_s$-meson and $\Lambda_s$-baryon spectra \cite{Witt} in the central
rapidity range $|\eta|<0.5$ at the energy $\sqrt s = 200$~GeV.
The spectra are presented in Figs. 10(a) and 11(a).
The multiplicity density was varied in the range $dN/d\eta=1.3-9.0$.
The transverse momentum distributions were measured up to
4.5~GeV/c.
The corresponding function $\psi(z)$ is shown in
Figs. 10(b) and 11(b), respectively.
The scaling behavior gives strong restriction
on the value of $c$. Data prefer $c=0.25$ in both cases.

Thus we conclude that the available experimental data on the multiplicity
dependence of spectra of charged hadrons, $K^0_s$-mesons  and
$\Lambda_s$-baryons produced in $pp$ collisions at RHIC confirm
generalized $z$-scaling for the same value of the
parameter $c$.

\vskip 0.5cm
{\section{Conclusions}}

Generalized $z$-scaling for the inclusive particle production in
proton-proton collisions was suggested. The scaling variable $z$
is function of multiplicity density $dN/d\eta|_0$ of charged
particles in the central region of collision. The variable $z$
depends on the parameters $c$, $\delta$ and $\epsilon$. They are
interpreted as specific heat of the produced medium, anomalous
fractal dimension of the proton and fractal dimension of the
fragmentation process, respectively. Connection between the
scaling variable $z$ and entropy $\it S$ of the interacting system
was established.

We have analyzed experimental data on inclusive cross sections of
hadrons ($h^{\pm}$, $\pi^{-}$, $K^{-}$, $K^0_S$, $\bar{p}$ and $\Lambda$)
measured in proton-proton collisions at FNAL,
ISR and RHIC. The data cover a wide range of collision energy,
transverse momenta and angles of the produced particles. Spectra from
minimum biased events and events with various multiplicity
selection criteria have been studied. The energy, angular and
multiplicity independence of the scaling function was established.
It gives strong constrains on the values of the parameters $c$,
$\delta$ and $\epsilon$. It was shown that the parameters are
constants in the considered kinematical region. The parameters
$c=0.25$ and $\delta=0.5$ were found to be the same for all types
of the considered inclusive hadrons. The value of $\epsilon$
increases with mass of the produced hadron.

The variable $z$ has property of a fractal measure connected with
parton content of the composite structures involved. The fractal
dimensions $\delta$ and $\epsilon$ determine fractal properties of
$z$ in space of the momentum fractions. The scaling function
$\psi(z)$ manifests two regimes of particle production. The hard
regime is characterized by the power law $\psi(z)\sim z^{-\beta}$
at large $z$. Soft processes correspond to  behavior of the
$\psi(z)$ at small $z$. Slope of the scaling curve decreases with
$z$ in this region. On basis of the performed analysis we conclude
that $z$-scaling in proton-proton collisions is regularity which
reflects self-similarity, locality and fractality
of hadron interaction at constituent level. It concerns structure
of the colliding objects, interactions of their constituents and
fragmentation process.

We consider that obtained results are of interest for searching
and study of new physics phenomena in particle production over a
wide range of collision energies, high transverse momenta and
large multiplicities in proton-proton and nucleus-nucleus
interactions at the RHIC and LHC.

\vskip 5mm
{\large \bf Acknowledgments.}
The investigations have been partially supported by the IRP
AVOZ10480505, by the Grant Agency of the Czech Republic under
the contract No. 202/04/0793 and by the special program of
the Ministry of Science and Education of the Russian Federation,
grant RNP.2.1.1.5409.

\vskip 0.5cm {\section{Appendix}}

The invariant differential cross section for production of the
inclusive particle is normalized as follows
\begin{equation}
\int E\frac{d^3\sigma}{dp^3} d\eta d^2p_{\bot} = \sigma_{inel}N.
\label{eq:r39}
\end{equation}
The $\sigma_{inel}$ is the inelastic cross section and $N$ is
the average multiplicity. The differential cross section of
identified hadrons of certain type is normalized to the
corresponding average multiplicity of this type. The inclusive
cross section can be expressed in terms of $\kappa_1$ and
$\kappa_2$ in the way
\begin{equation}
E\frac{d^3\sigma}{dp^3} = -\frac{1}{2\pi}
\frac{\sqrt{(P_1P_2)^2-M_1^2M_2^2}}{[(P_1P_2)-M_1M_2]^2}
\frac{d^2\sigma}{d\kappa_1 d\kappa_2}.
\label{eq:r40}
\end{equation}
In the region of high energies, the formula can be written in the
approximate form
\begin{equation}
E\frac{d^3\sigma}{dp^3} = -\frac{1}{\pi s}
\frac{d^2\sigma}{d\kappa_1 d\kappa_2},
\label{eq:r41}
\end{equation}
where $s$ is square of the centre-of-mass energy. We suppose that
the inclusive cross section is given by solution (\ref{eq:r6}) as
function of a single variable $z =z(\kappa_1,\kappa_2)$. Another
independent combination of $\kappa_1$ and $\kappa_2$ is rapidity
\begin{equation}
y=\frac{1}{2}\ln \frac{\kappa_2}{\kappa_1}.
\label{eq:r42}
\end{equation}
Using the variables $z$ and $y$, we get the normalization
\begin{equation}
\int \frac{d^2\sigma}{d\kappa_1 d\kappa_2} d\kappa_1 d\kappa_2 =
\int \frac{d^2\sigma}{dy dz} dy dz = \sigma_{inel}\int
\rho(y)\psi(z) dy dz = \sigma_{inel}N,
\label{eq:r43}
\end{equation}
where $\rho(y)\equiv dN/dy$ is the rapidity distribution
of particles of the considered type.  Detailed analysis of
experimental data on angular properties of the scaling function
showed that the factorization
\begin{equation}
\frac{d^2\sigma}{d\eta dz}  = \sigma_{inel} \rho(\eta)\psi(z)
\label{eq:r44}
\end{equation}
is valid using rather pseudorapidity than rapidity. In the
centre-of-mass of the symmetric systems (e.g. for $NN$ collisions)
we can exploit the relations
\begin{equation}
\eta= \frac{1}{2}\ln \frac{1-cos\theta_2}{1+cos\theta_1} \sim
y=\frac{1}{2}\ln \frac{\kappa_2}{\kappa_1},
\label{eq:r45}
\end{equation}
\begin{equation}
\kappa_1 = \frac{(P_2p)}{(P_1P_2)-M_1M_2} \sim
\frac{E_p+p\cos\theta_1}{\sqrt{s}}, \ \ \ \ \kappa_2 =
\frac{(P_1p)}{(P_1P_2)-M_1M_2} \sim \frac{E_p-p\cos\theta_2}{\sqrt{s}}
\label{eq:r46}
\end{equation}
and set at the end $\theta_1=\theta_2\equiv\theta$. Then we get
\begin{equation}
\frac{\partial \eta }{\partial\kappa_1}= \frac{\partial
\eta}{\partial\theta_1} \frac{\partial\theta_1}{\partial\kappa_1}
= -\frac{\sqrt{s}}{2p}\frac{1}{(1\!+\!cos\theta)},
\label{eq:r47}
\end{equation}
\begin{equation}
\frac{\partial \eta }{\partial\kappa_2}= \frac{\partial
\eta}{\partial\theta_2} \frac{\partial\theta_2}{\partial\kappa_2}
= +\frac{\sqrt{s}}{2p}\frac{1}{(1\!-\!cos\theta)}.
\label{eq:r48}
\end{equation}
The derivatives $\partial z/\partial\kappa_i$ which enter the
Jacobian (\ref{eq:r34}) were calculated numerically.

Finally, we examine the transverse kinetic energy
$s^{1/2}_{\bot}$ which enter the definition (\ref{eq:r22}) of the scaling variable $z$.
The transverse momentum balance in the
constituent sub-process is guarantied by the identity
\begin{equation}
\chi_1\chi_2=\mu_1\mu_2=\lambda_1\lambda_2+\lambda_0.
\label{eq:r49}
\end{equation}
Neglecting masses $M_i\lambda_i$ and $M_i\chi_i$ of the interacting constituents in the expressions
(\ref{eq:r24}) and (\ref{eq:r25}), we get  for the transverse kinetic
energies $T_a$ and $T_b$ the  following relations
\begin{equation}
T_{\bot}^a \sim y_a\sqrt{\lambda_1\lambda_2 2P_1P_2}-m_1 \sim
\sqrt{(p_{\bot}^a)^2+m_1^2}-m_1
\label{eq:r50}
\end{equation}
and
\begin{eqnarray}
T_{\bot}^b \sim y_b\sqrt{\chi_1\chi_2 2P_1P_2}-m_2 \sim
y_b\sqrt{(\lambda_1\lambda_2+\lambda_0)s}-m_2
\nonumber \\
\sim
y_b\sqrt{\kappa_1\kappa_2s/y_a^2+m_2^2/y_b^2-m_1^2/y_a^2}-m_2 \nonumber \\
\sim y_b\sqrt{((p_{\bot}^a)^2+m_1^2)/y_a^2
+m_2^2/y_b^2-m_1^2/y_a^2}-m_2 = \sqrt{(p^b_{\bot})^2+m_2^2} - m_2.
 \label{eq:r51}
\end{eqnarray}
In the last equation we have used the transverse momentum balance
\begin{equation}
p_{\bot}^a/y_a = p_{\bot}^b/y_b.
\label{eq:r52}
\end{equation}
valid in the constituent sub-process. The $p^a$ and $p^b$ are
momenta of the inclusive particle ($m_1$) and its
counterpart ($m_2$), respectively.

\newpage
\begin{minipage}{4cm}

\vskip 3cm
\end{minipage}

\vskip 9cm
\begin{center}
\parbox{7cm}{\epsfxsize=7.cm\epsfysize=7.cm\epsfbox[95 95 400 400]
{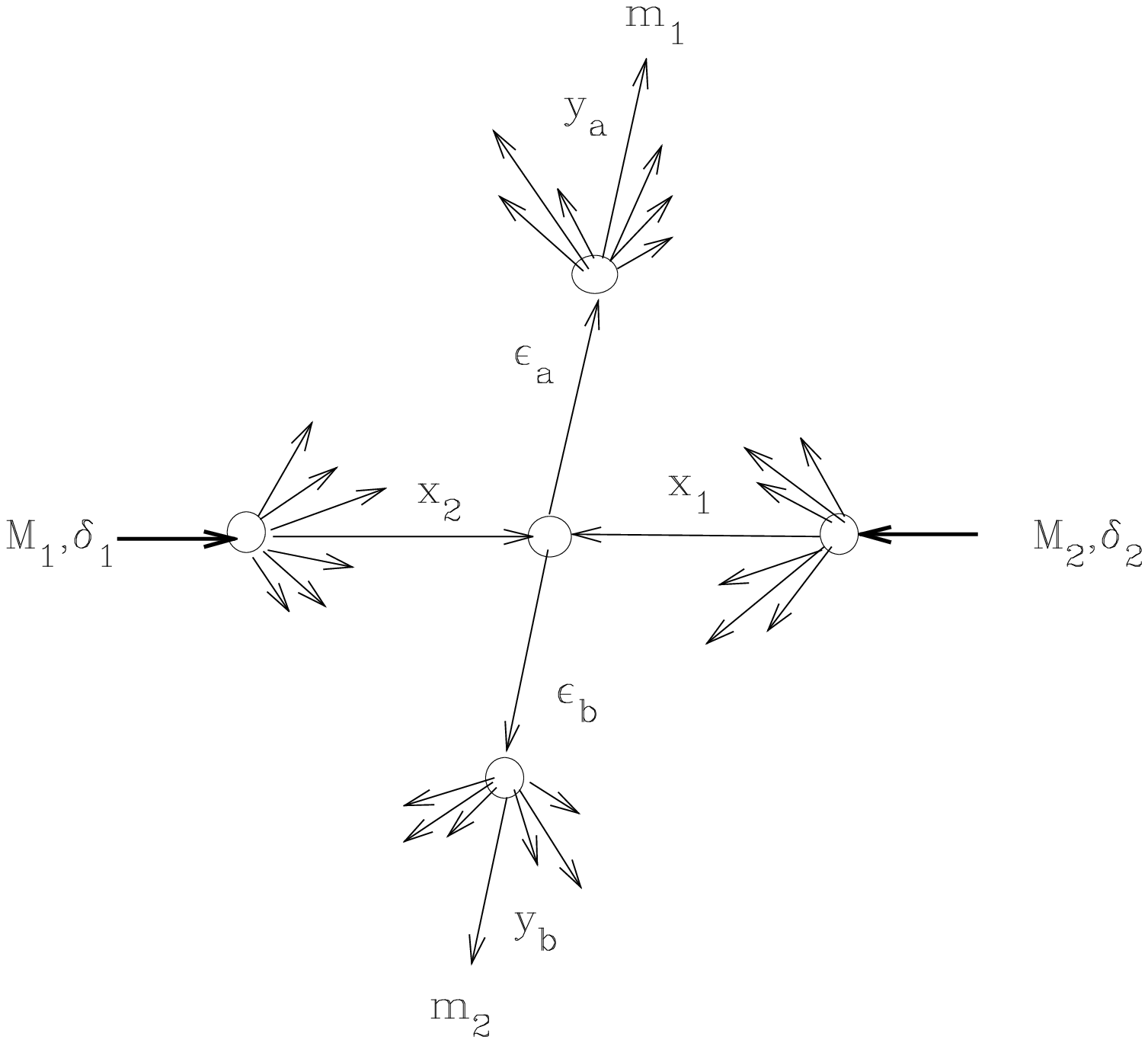}{}}
\vskip 0.cm

{\bf Figure 1.} Diagram of the constituent sub-process.
\end{center}

\newpage
\begin{minipage}{4cm}

\end{minipage}

\vskip 4cm
\begin{center}
\hspace*{-1.5cm}
\parbox{6cm}{\epsfxsize=6.cm\epsfysize=6.cm\epsfbox[95 95 400 400]
{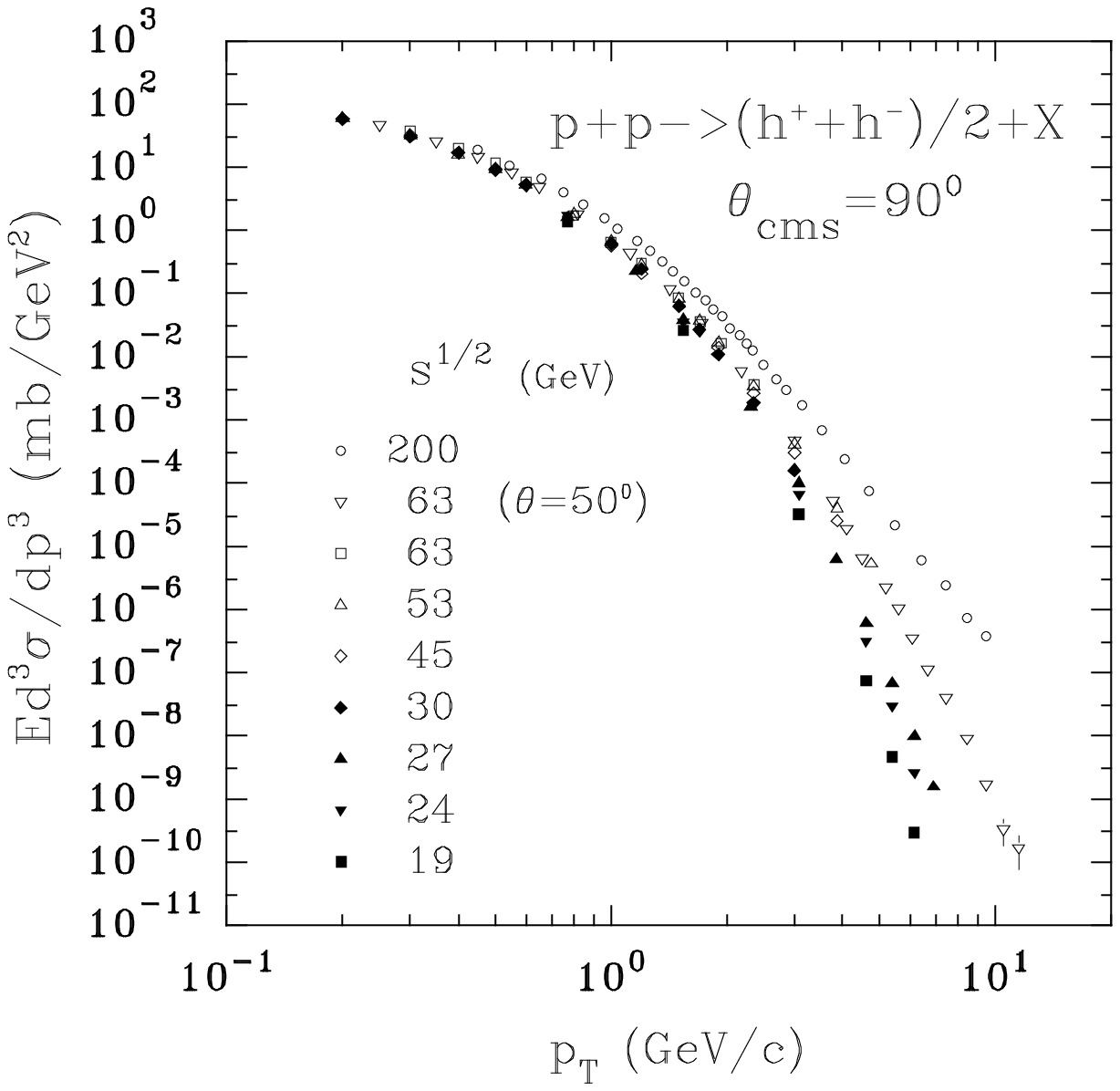}{}}
\hspace*{2cm}
\parbox{6cm}{\epsfxsize=6.cm\epsfysize=6.cm\epsfbox[95 95 400 400]
{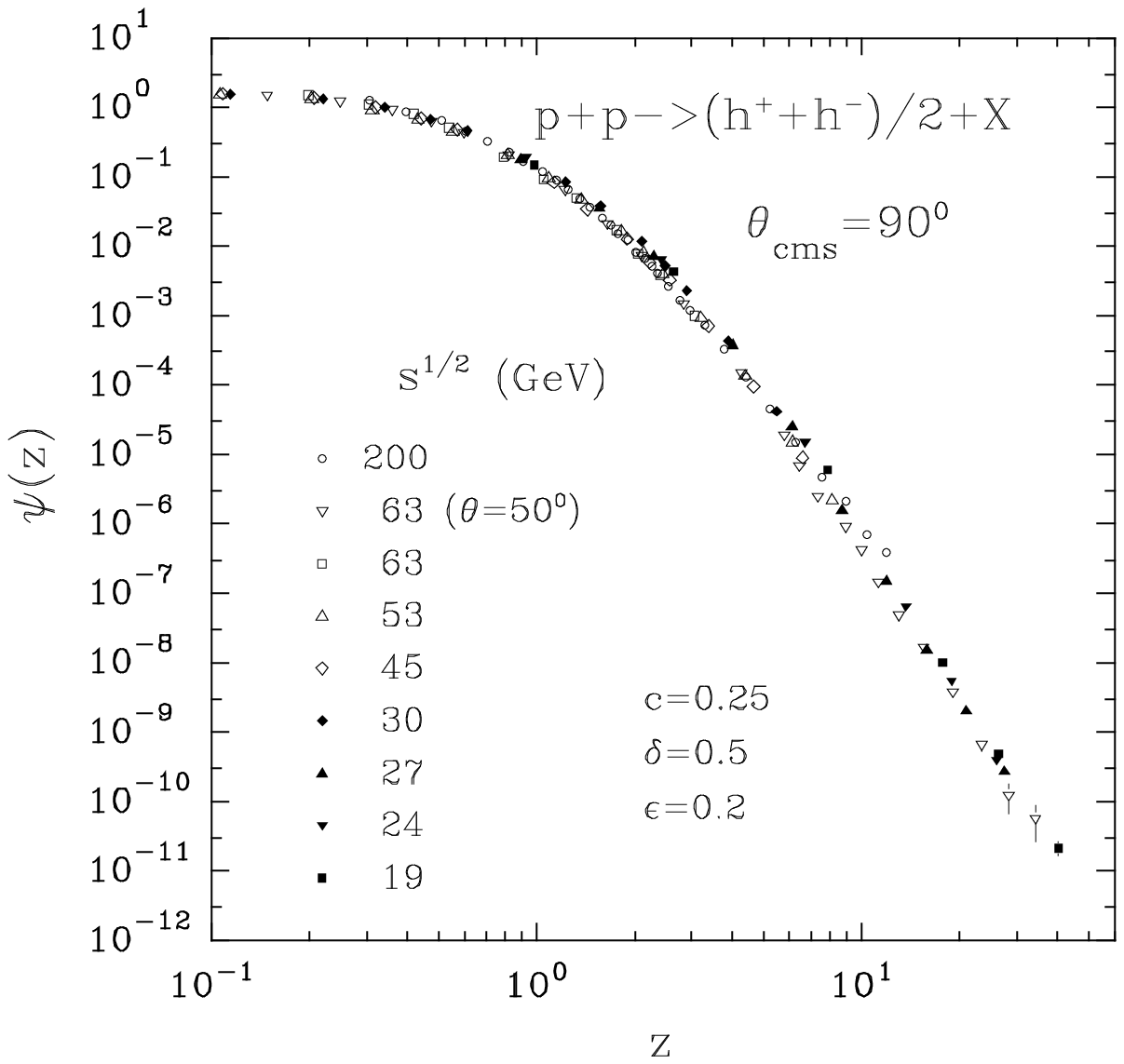}{}}
\vskip -2.cm
\hspace*{0.cm} a) \hspace*{8.cm} b)\\[0.5cm]
\end{center}

{\bf Figure 2.} (a) Transverse momentum spectra of charged hadrons produced
in $pp$ collisions  at $\sqrt s=19-200$~GeV.
Experimental data are taken from Refs. \cite{FNAL1,ISR,CDHW,STAR1}.
(b) The corresponding scaling function $\psi(z)$.

\vskip 5cm

\begin{center}
\hspace*{-1.5cm}
\parbox{6cm}{\epsfxsize=6.cm\epsfysize=6.cm\epsfbox[95 95 400 400]
{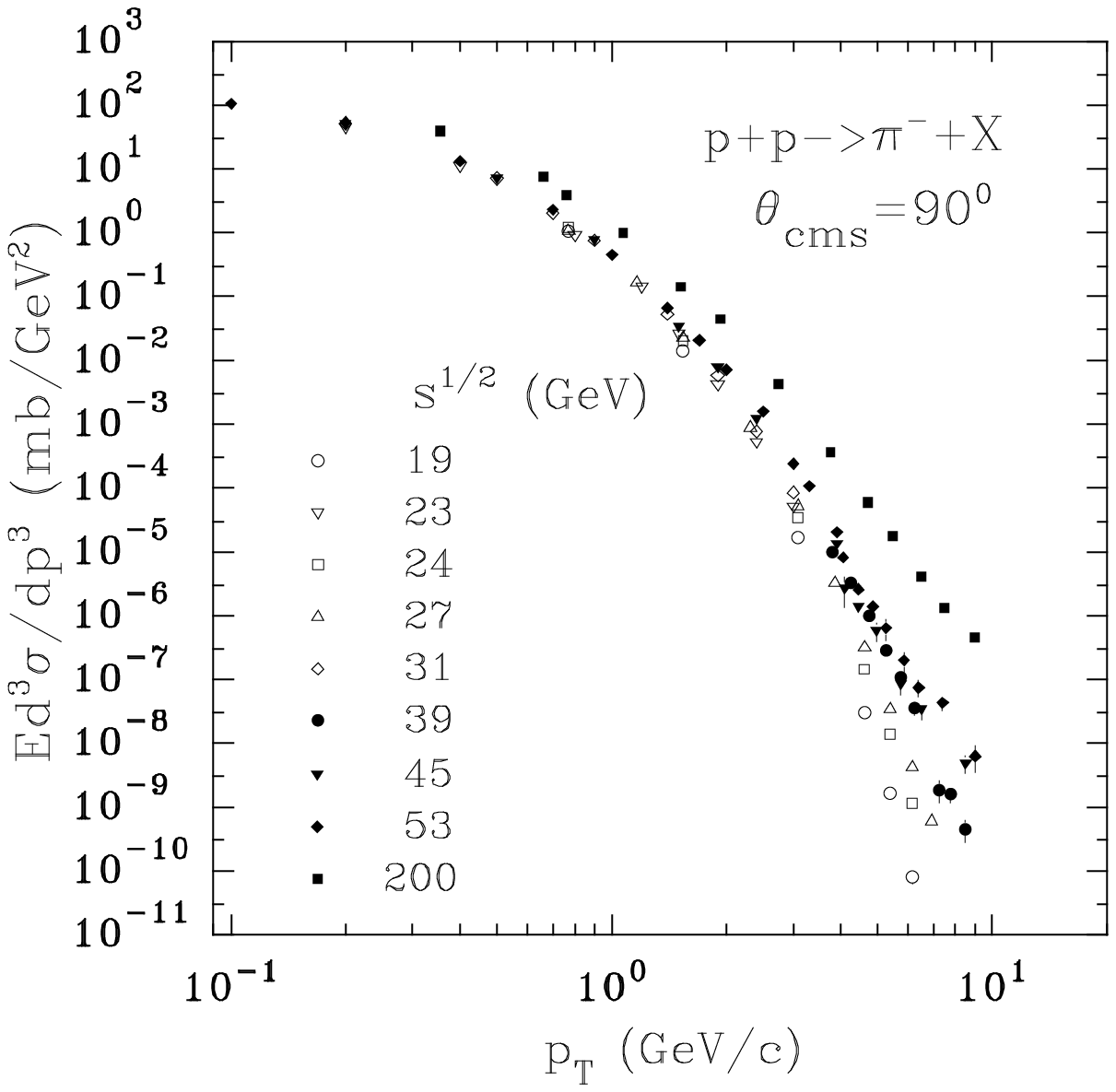}{}}
\hspace*{2cm}
\parbox{6cm}{\epsfxsize=6.cm\epsfysize=6.cm\epsfbox[95 95 400 400]
{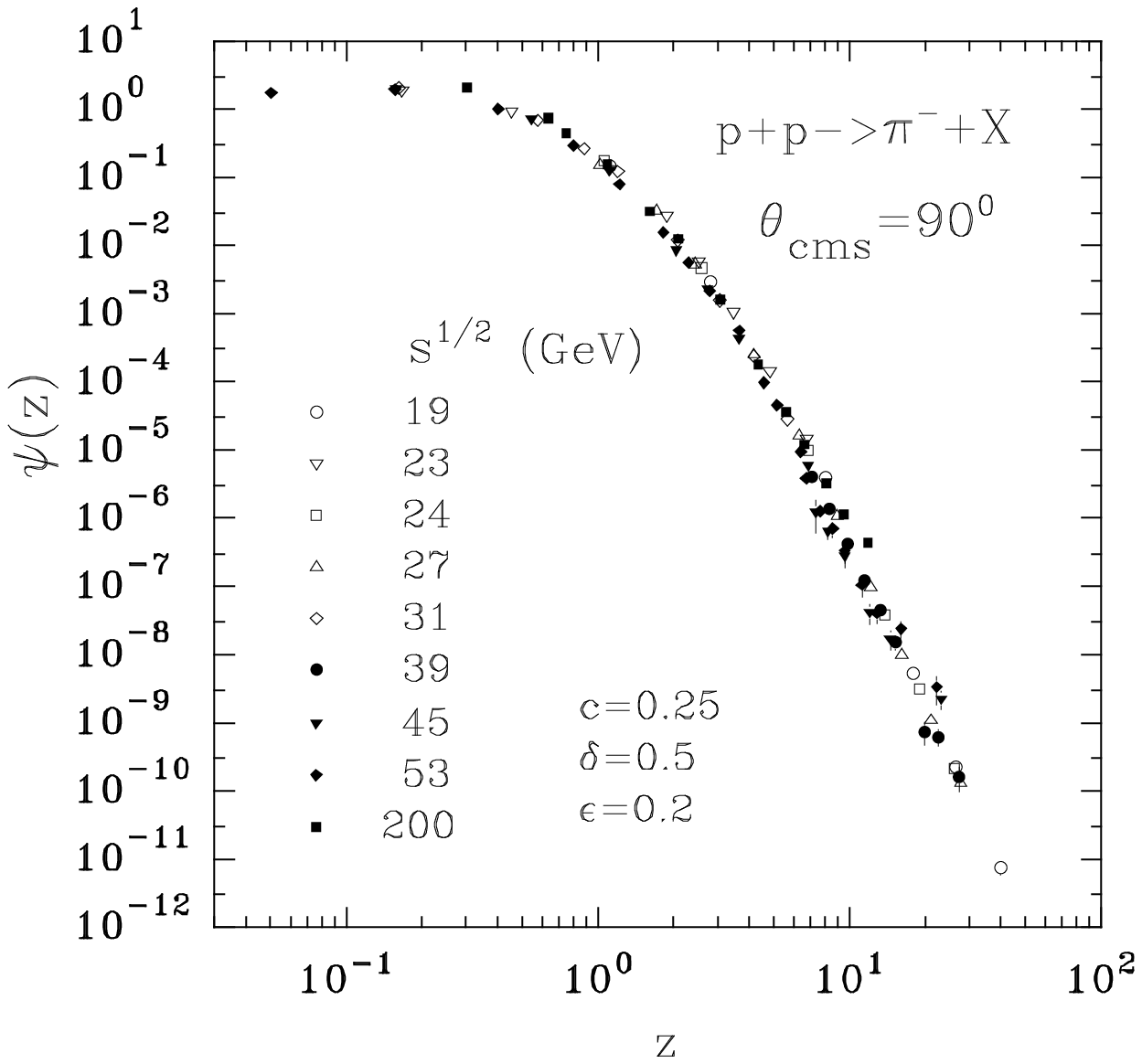}{}}
\vskip -2.cm
\hspace*{0.cm} a) \hspace*{8.cm} b)\\[0.5cm]
\end{center}

{\bf Figure 3.}
a) Transverse momentum spectra of $\pi^-$-mesons produced
in $pp$ collisions at $\sqrt s=19-200$~GeV.
Experimental data are taken from Refs. \cite{FNAL1,ISR,FNAL2,Baranikova}.
(b) The corresponding scaling function $\psi(z)$.

\newpage
\begin{minipage}{4cm}

\end{minipage}

\vskip 4cm
\begin{center}
\hspace*{-1.5cm}
\parbox{6cm}{\epsfxsize=6.cm\epsfysize=6.cm\epsfbox[95 95 400 400]
{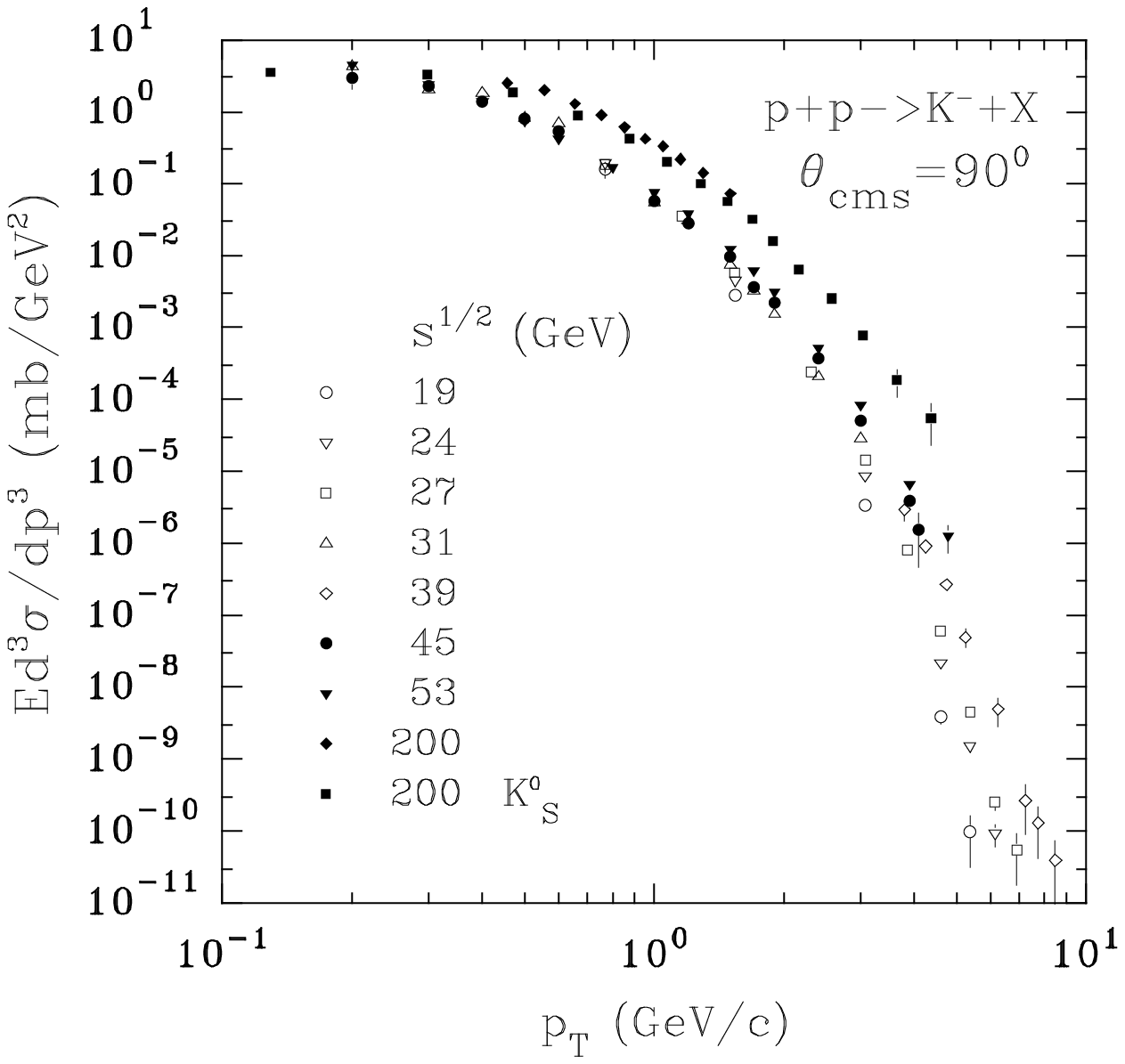}{}}
\hspace*{2.cm}
\parbox{6cm}{\epsfxsize=6.cm\epsfysize=6.cm\epsfbox[95 95 400 400]
{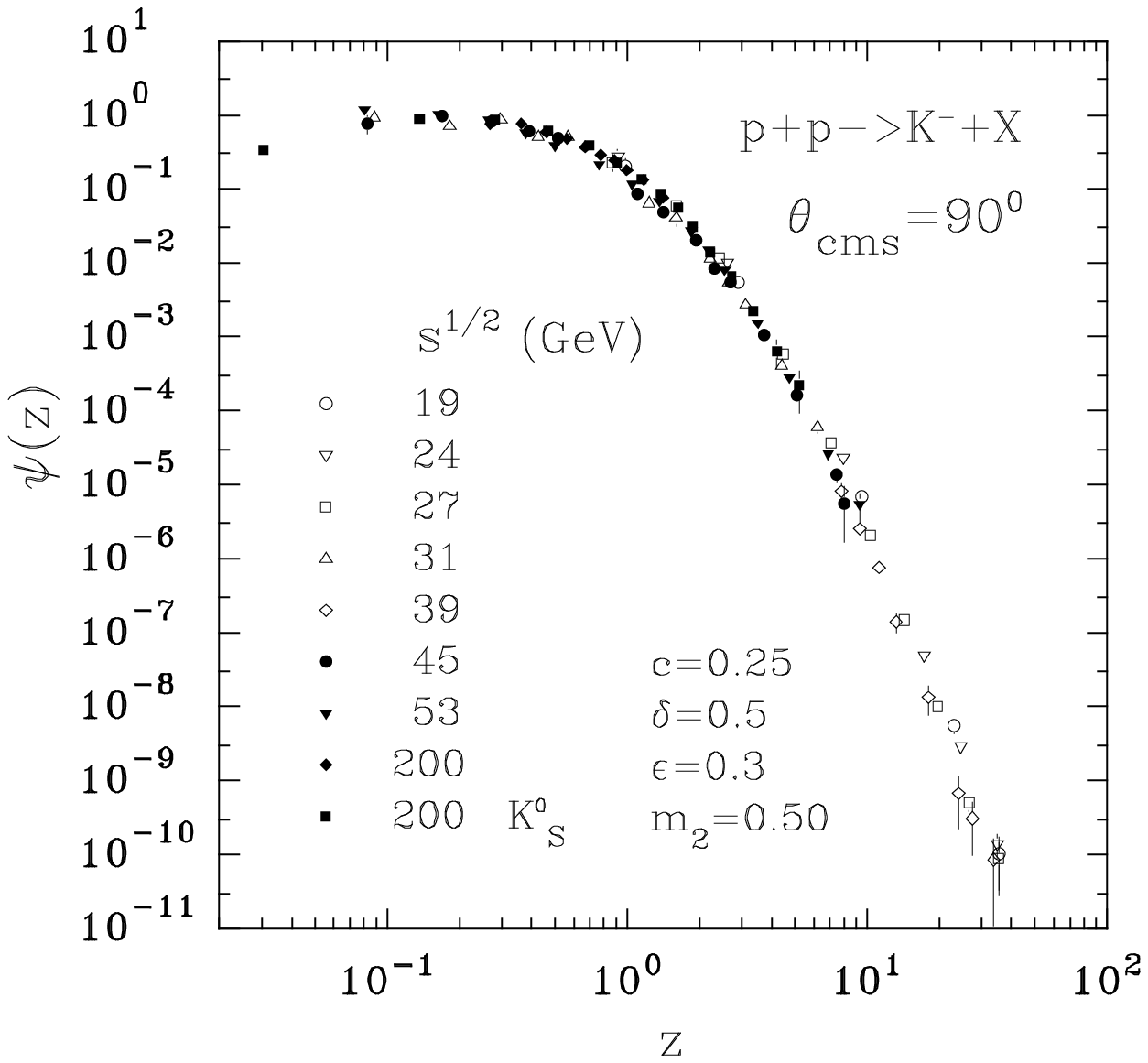}{}}
\vskip -2cm
\hspace*{0.1cm} a) \hspace*{8.cm} b)\\[0.5cm]
\end{center}

{\bf Figure 4.}
(a) Transverse momentum spectra of $K^-$-mesons produced
in $ pp$ collisions at $\sqrt s=19-200$~GeV.
The spectrum of $K^0_s$-mesons is shown by full squares.
Experimental data are taken from Refs. \cite{FNAL1,ISR,FNAL2,STAR2,Adams}.
(b) The corresponding scaling function $\psi(z)$.

\vskip 5cm

\begin{center}
\hspace*{-1.5cm}
\parbox{6cm}{\epsfxsize=6.cm\epsfysize=6.cm\epsfbox[95 95 400 400]
{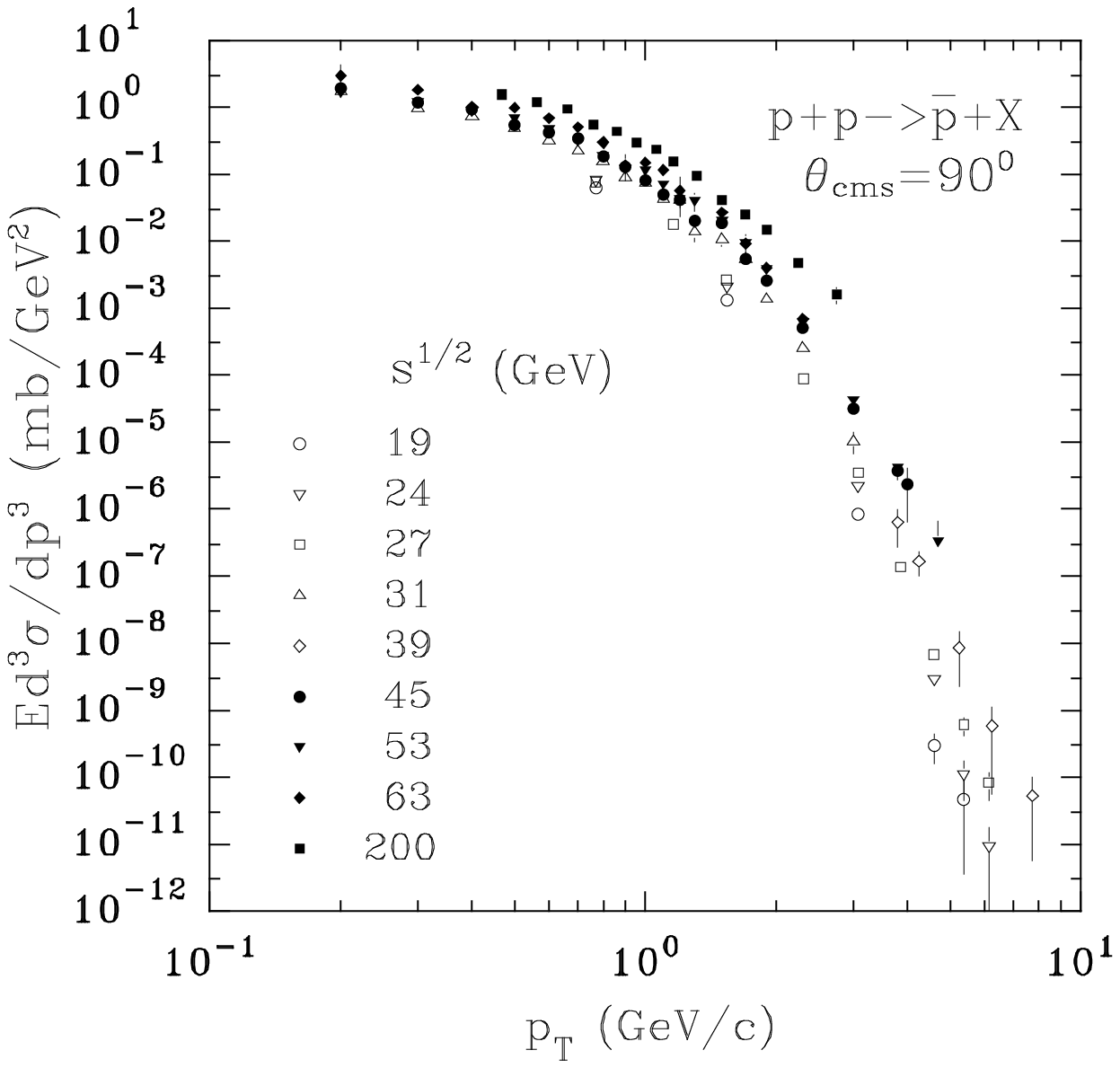}{}}
\hspace*{2.cm}
\parbox{6cm}{\epsfxsize=6.cm\epsfysize=6.cm\epsfbox[95 95 400 400]
{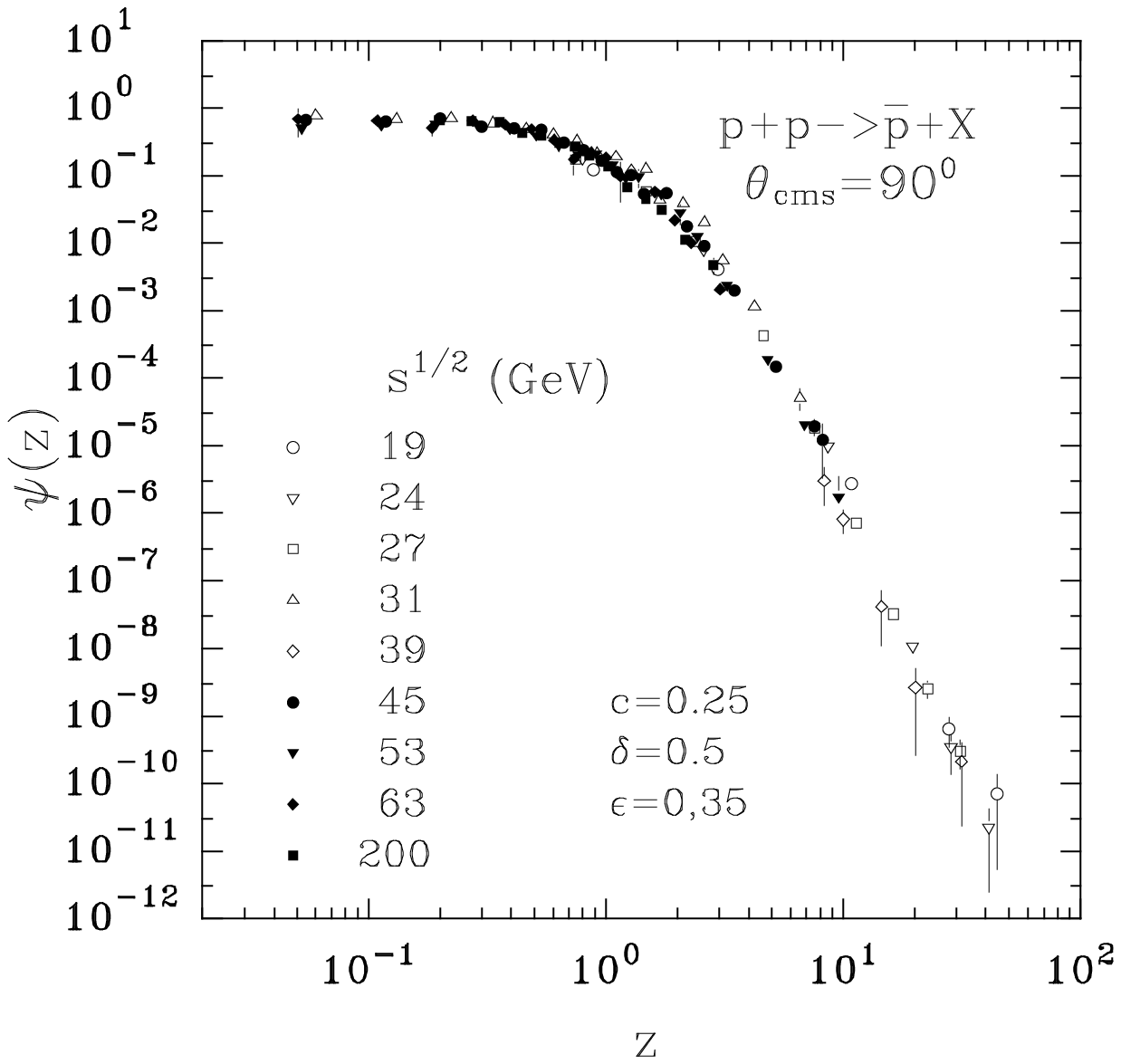}{}}
\vskip -2.cm
\hspace*{0.cm} a) \hspace*{8.cm} b)\\[0.5cm]
\end{center}

{\bf Figure 5.}
(a) Transverse momentum spectra of antiprotons produced in  $pp$ collisions
at $\sqrt s=19-200$~GeV.
Experimental data are taken from Refs. \cite{FNAL1,ISR,FNAL2,STAR2} .
(b) The corresponding scaling function $\psi(z)$.

\newpage
\begin{minipage}{4cm}

\end{minipage}

\vskip 4cm
\begin{center}
\hspace*{-1.5cm}
\parbox{6cm}{\epsfxsize=6.cm\epsfysize=6.cm\epsfbox[95 95 400 400]
{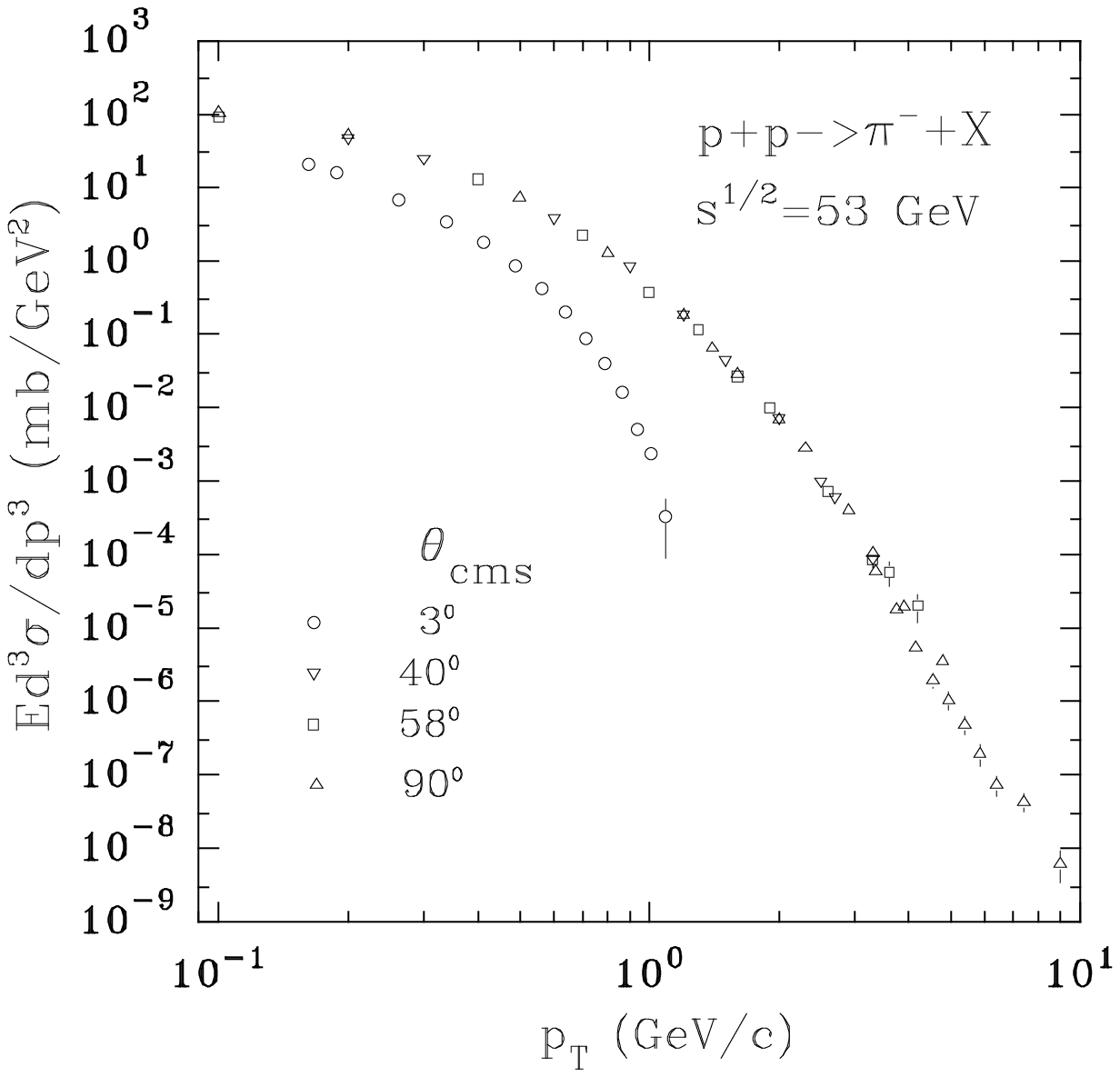}{}}
\hspace*{2.cm}
\parbox{6cm}{\epsfxsize=6.cm\epsfysize=6.cm\epsfbox[95 95 400 400]
{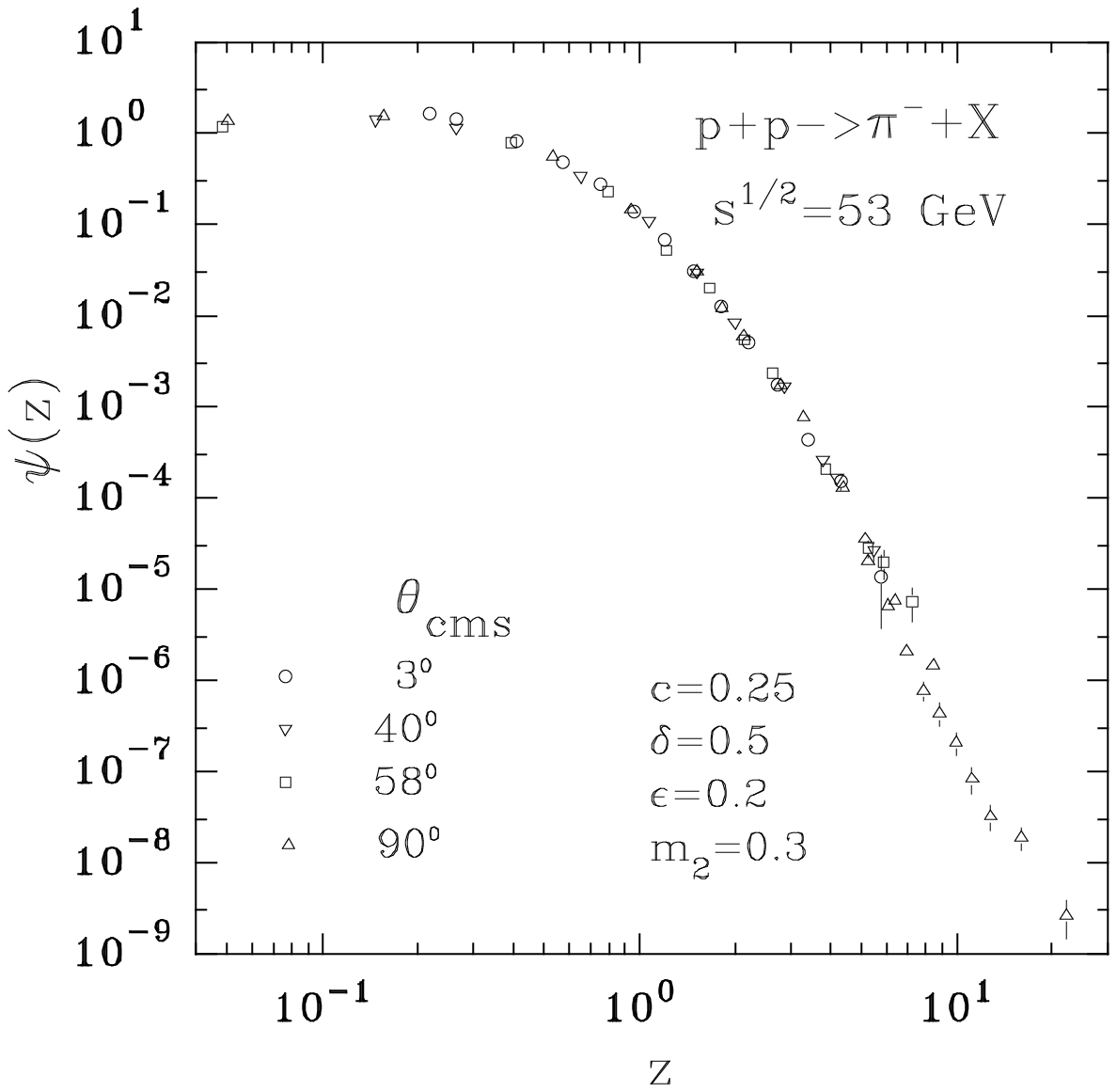}{}}
\vskip -2cm
\hspace*{0.1cm} a) \hspace*{8.cm} b)\\[0.5cm]
\end{center}

{\bf Figure 6.} (a) Transverse momentum spectra of $\pi^-$-mesons
produced in $ pp$ collisions for different angles at $\sqrt
s=53$~GeV. Experimental data are taken from Refs. \cite{ISR,CHLM}.
(b) The corresponding scaling function $\psi(z)$.

\vskip 5cm

\begin{center}
\hspace*{-1.5cm}
\parbox{6cm}{\epsfxsize=6.cm\epsfysize=6.cm\epsfbox[95 95 400 400]
{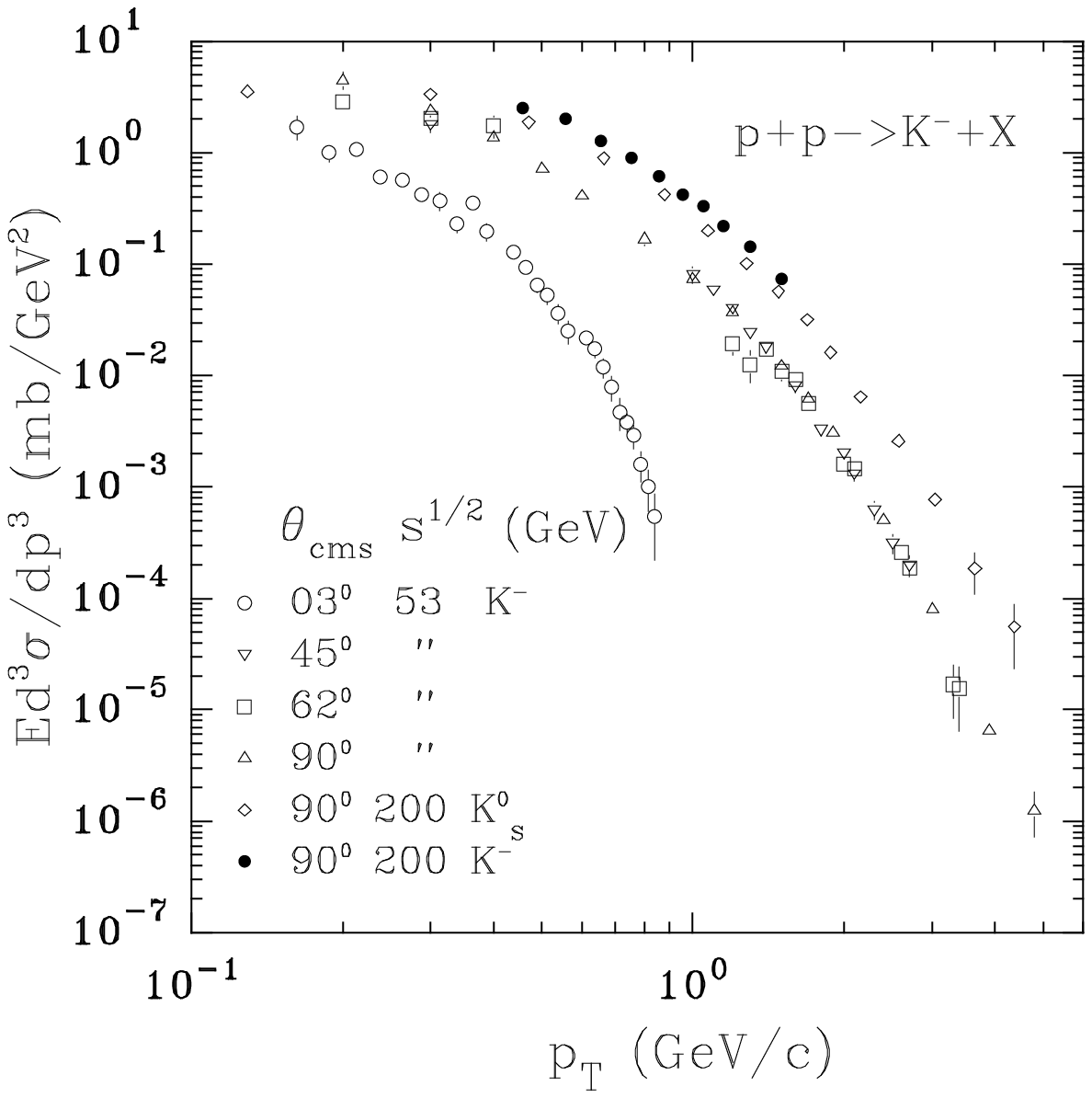}{}}
\hspace*{2.cm}
\parbox{6cm}{\epsfxsize=6.cm\epsfysize=6.cm\epsfbox[95 95 400 400]
{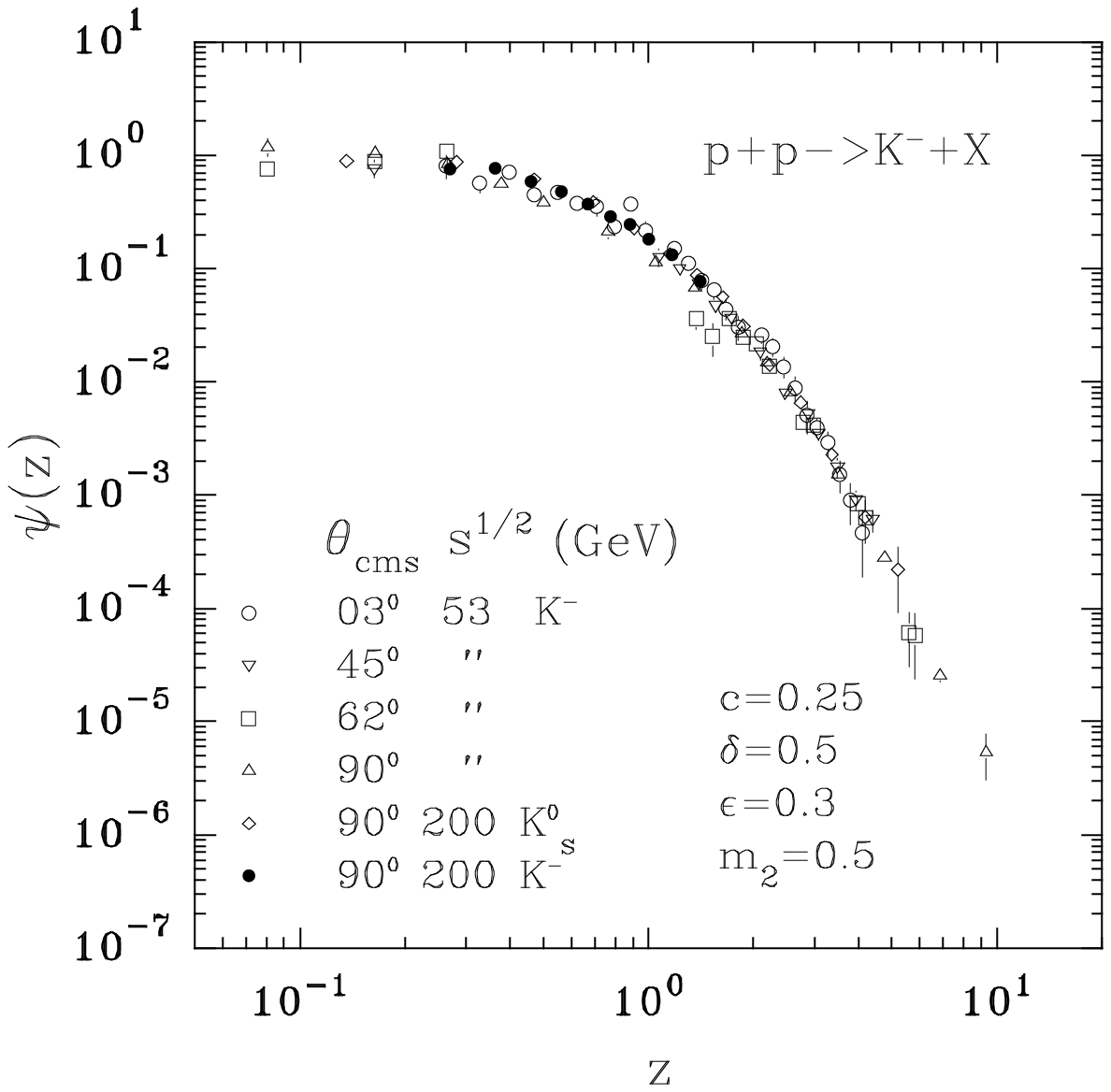}{}}
\vskip -2.cm
\hspace*{0.cm} a) \hspace*{8.cm} b)\\[0.5cm]
\end{center}

{\bf Figure 7.}
(a) Transverse momentum spectra of $K^-$-mesons produced
in $ pp$ collisions for different angles at $\sqrt s=53$~GeV.
The $K^-$ and $K^0_s$-meson spectra for $\theta_{cms}\simeq 90$ at $\sqrt s=200$~GeV
are shown by $\bullet$ and $\diamond$, respectively.
Experimental data are taken from Refs. \cite{ISR,STAR2,Adams,CHLM}.
(b) The corresponding scaling function $\psi(z)$.

\newpage
\begin{minipage}{4cm}

\end{minipage}

\vskip 4cm
\begin{center}
\hspace*{-1.5cm}
\parbox{6cm}{\epsfxsize=6.cm\epsfysize=6.cm\epsfbox[95 95 400 400]
{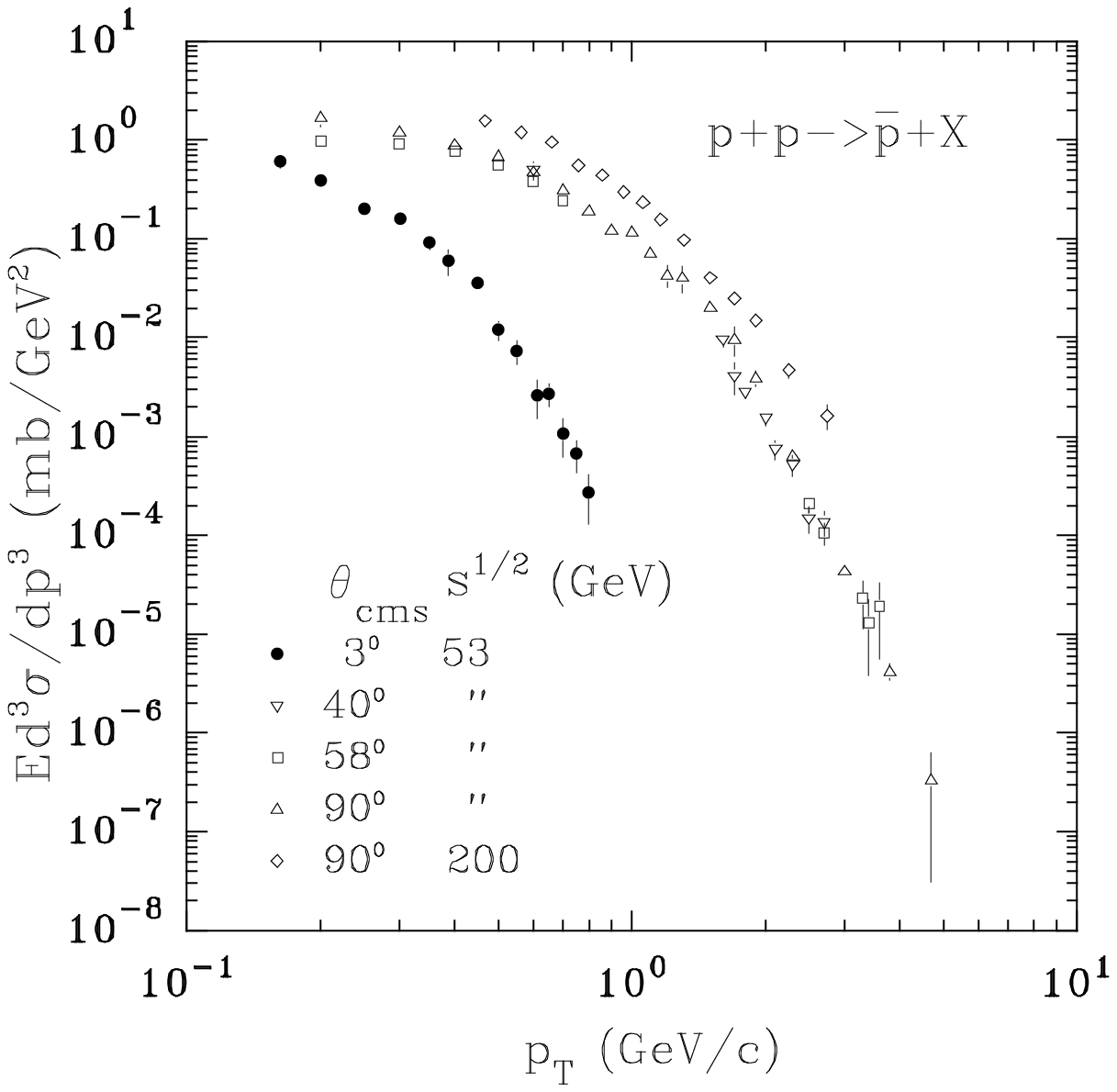}{}}
\hspace*{2.cm}
\parbox{6cm}{\epsfxsize=6.cm\epsfysize=6.cm\epsfbox[95 95 400 400]
{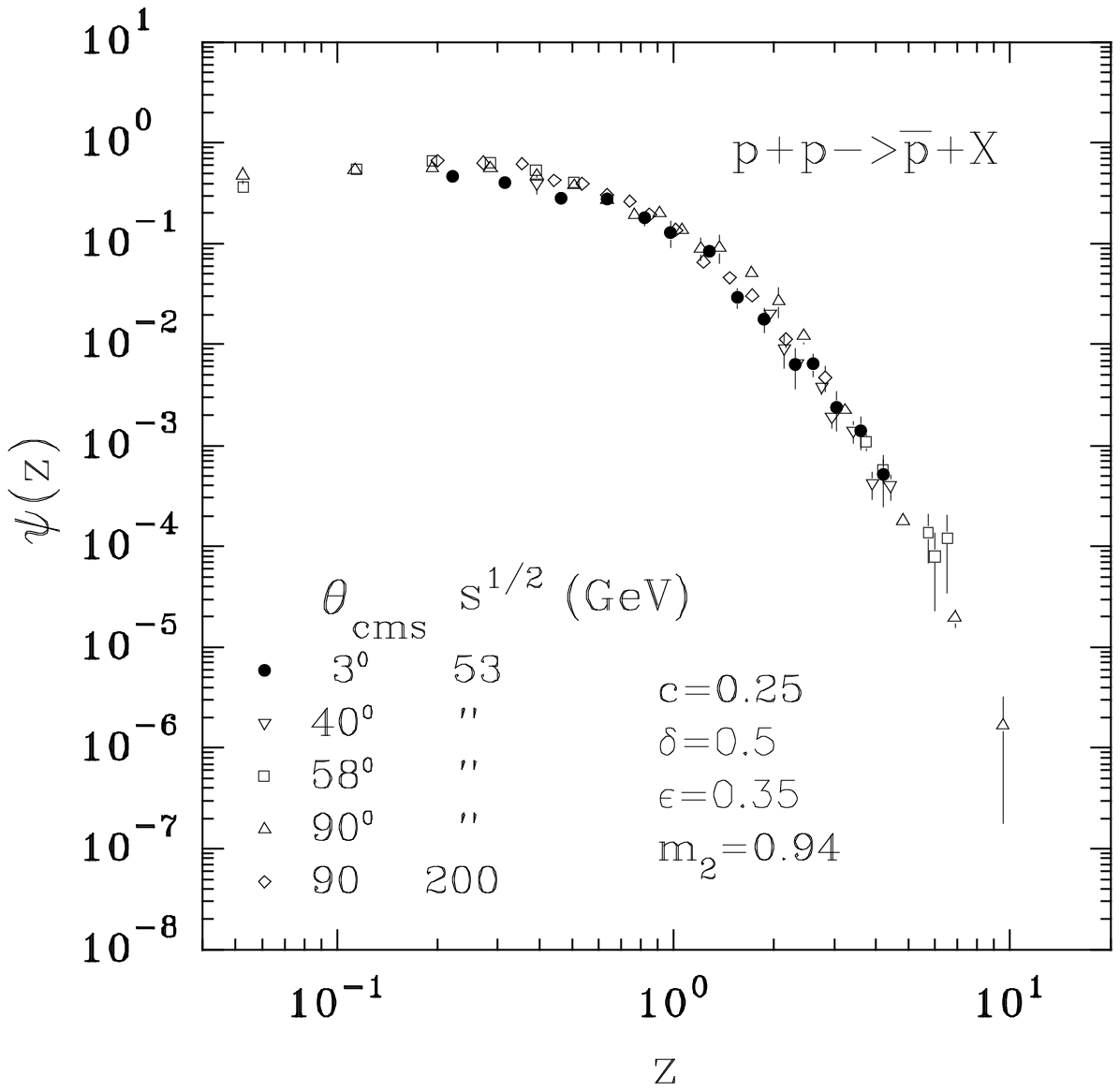}{}}
\vskip -2cm
\hspace*{0.1cm} a) \hspace*{8.cm} b)\\[0.5cm]
\end{center}

{\bf Figure 8.}
(a) Transverse momentum spectra of antiprotons produced
in $ pp$ collisions for different angles at $\sqrt s=53$~GeV.
The antiproton spectrum for $\theta_{cms}\simeq 90$ at $\sqrt s=200$~GeV
are shown by $\diamond$.
Experimental data are taken from Refs. \cite{ISR,STAR2,CHLM}.
(b) The corresponding scaling function $\psi(z)$.

\vskip 5cm

\begin{center}
\hspace*{-1.5cm}
\parbox{6cm}{\epsfxsize=6.cm\epsfysize=6.cm\epsfbox[95 95 400 400]
{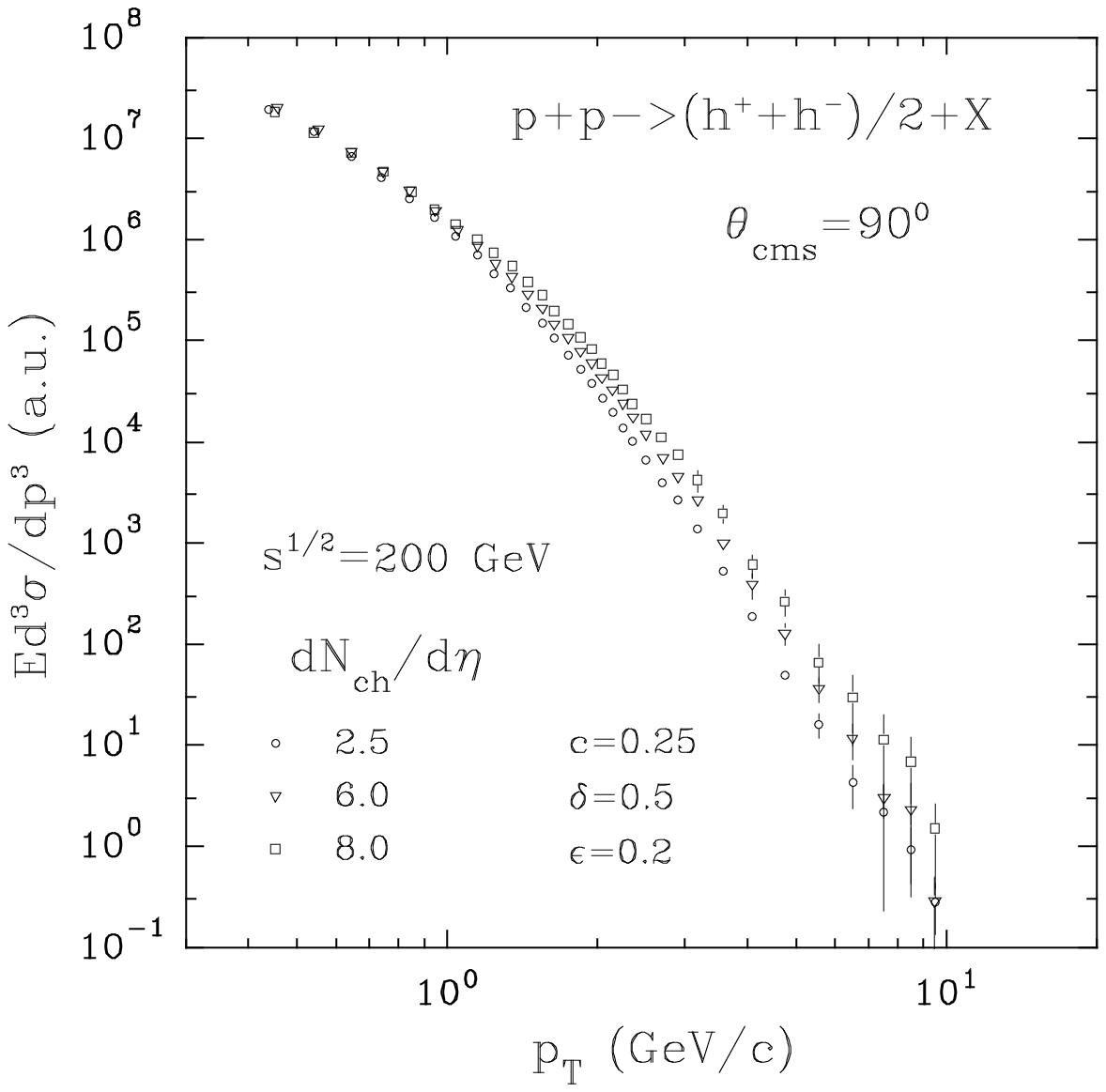}{}}
\hspace*{2.cm}
\parbox{6cm}{\epsfxsize=6.cm\epsfysize=6.cm\epsfbox[95 95 400 400]
{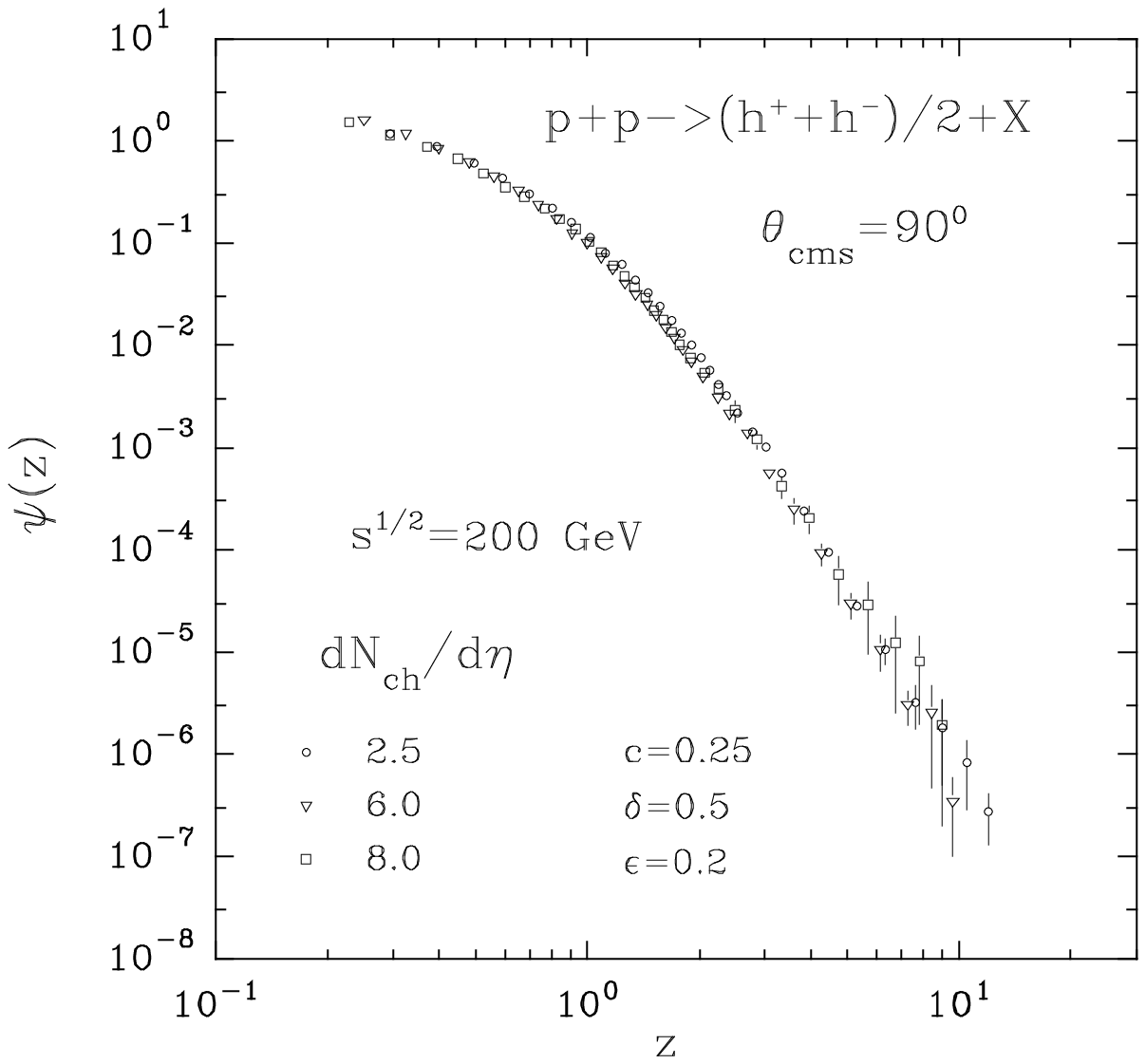}{}}
\vskip -2.cm
\hspace*{0.cm} a) \hspace*{8.cm} b)\\[0.5cm]
\end{center}

{\bf Figure 9.}
(a) Transverse momentum spectra of charged hadrons produced
in $ pp$ collisions for different multiplicity densities at $\sqrt s=200$~GeV.
The spectra are normalized at $p_T=0.4$~GeV/c.
Experimental data are taken from Ref. \cite{Gans}.
(b) The corresponding scaling function $\psi(z)$.

\newpage
\begin{minipage}{4cm}

\end{minipage}

\vskip 4cm
\begin{center}
\hspace*{-1.5cm}
\parbox{6cm}{\epsfxsize=6.cm\epsfysize=6.cm\epsfbox[95 95 400 400]
{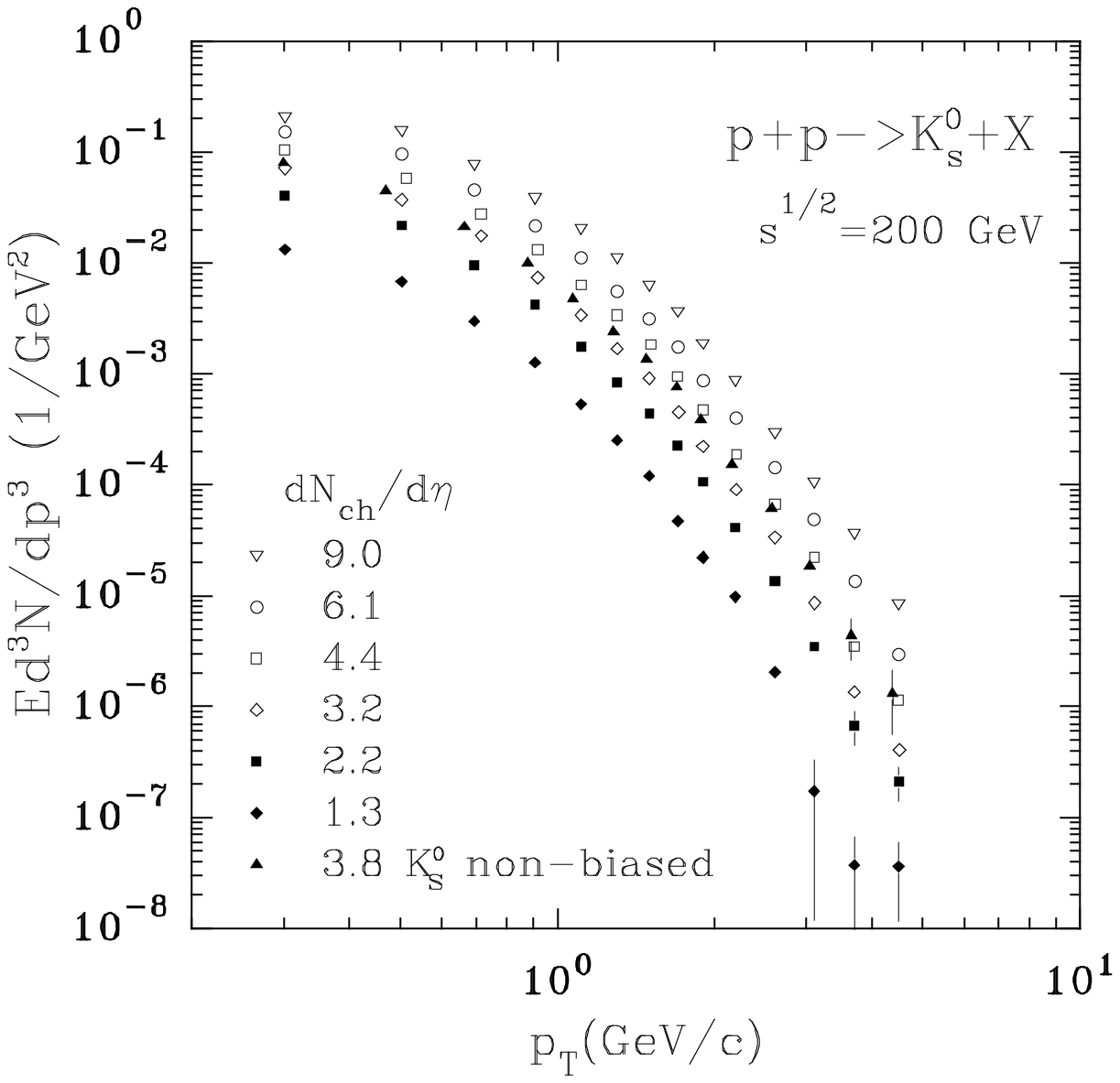}{}}
\hspace*{2.cm}
\parbox{6cm}{\epsfxsize=6.cm\epsfysize=6.cm\epsfbox[95 95 400 400]
{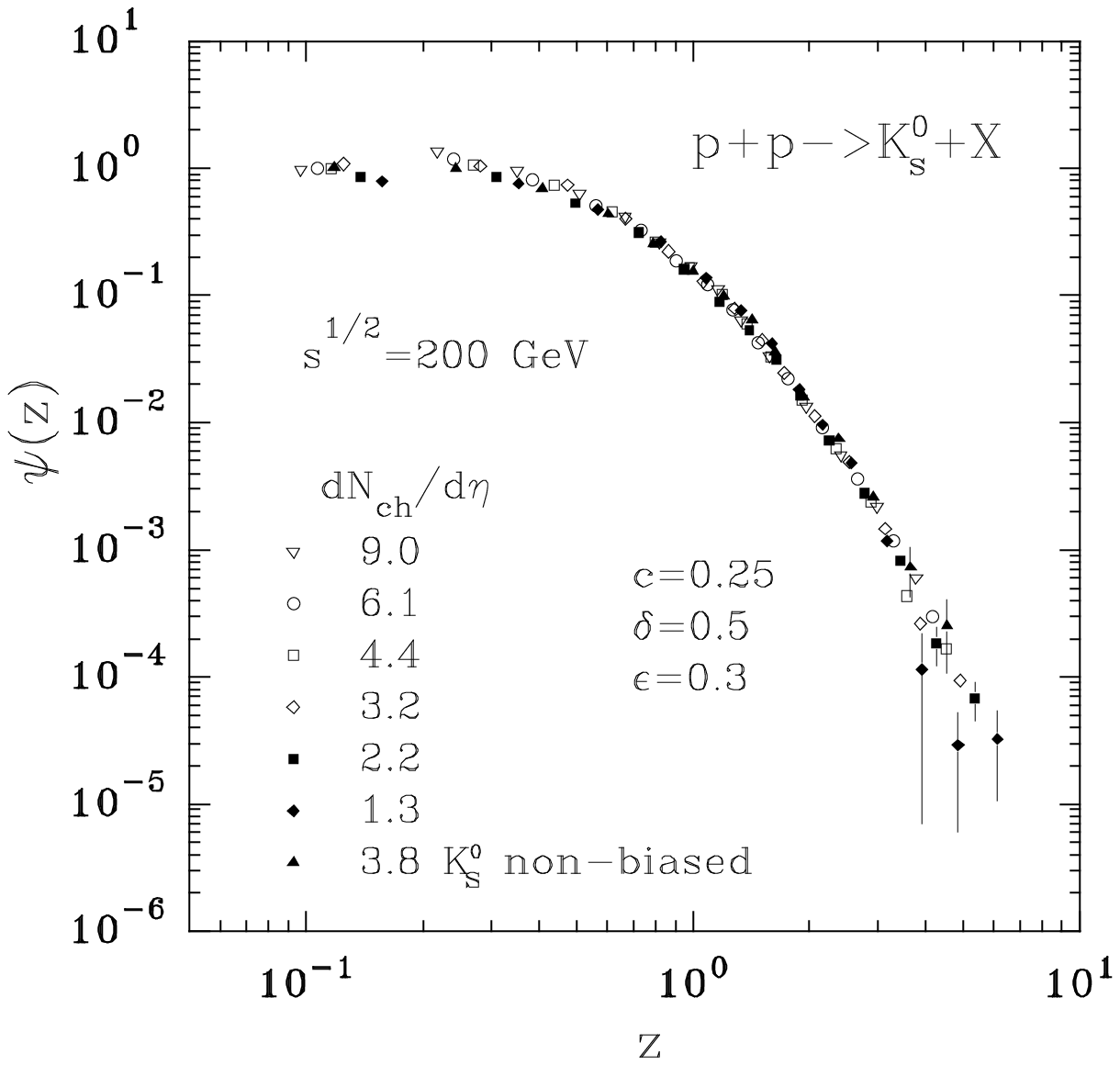}{}}
\vskip -2cm
\hspace*{0.1cm} a) \hspace*{8.cm} b)\\[0.5cm]
\end{center}

{\bf Figure 10.}
(a) Transverse momentum spectra of $K^0_s$-mesons produced
in $pp$ collisions for different multiplicity densities at $\sqrt s=200$~GeV.
The $K^0_s$-meson spectrum for non-biased $pp$ collisions at $\sqrt s=200$~GeV
is shown by black triangles.
Experimental data are taken from Ref. \cite{Witt}.
(b) The corresponding scaling function $\psi(z)$.

\vskip 5cm

\begin{center}
\hspace*{-1.5cm}
\parbox{6cm}{\epsfxsize=6.cm\epsfysize=6.cm\epsfbox[95 95 400 400]
{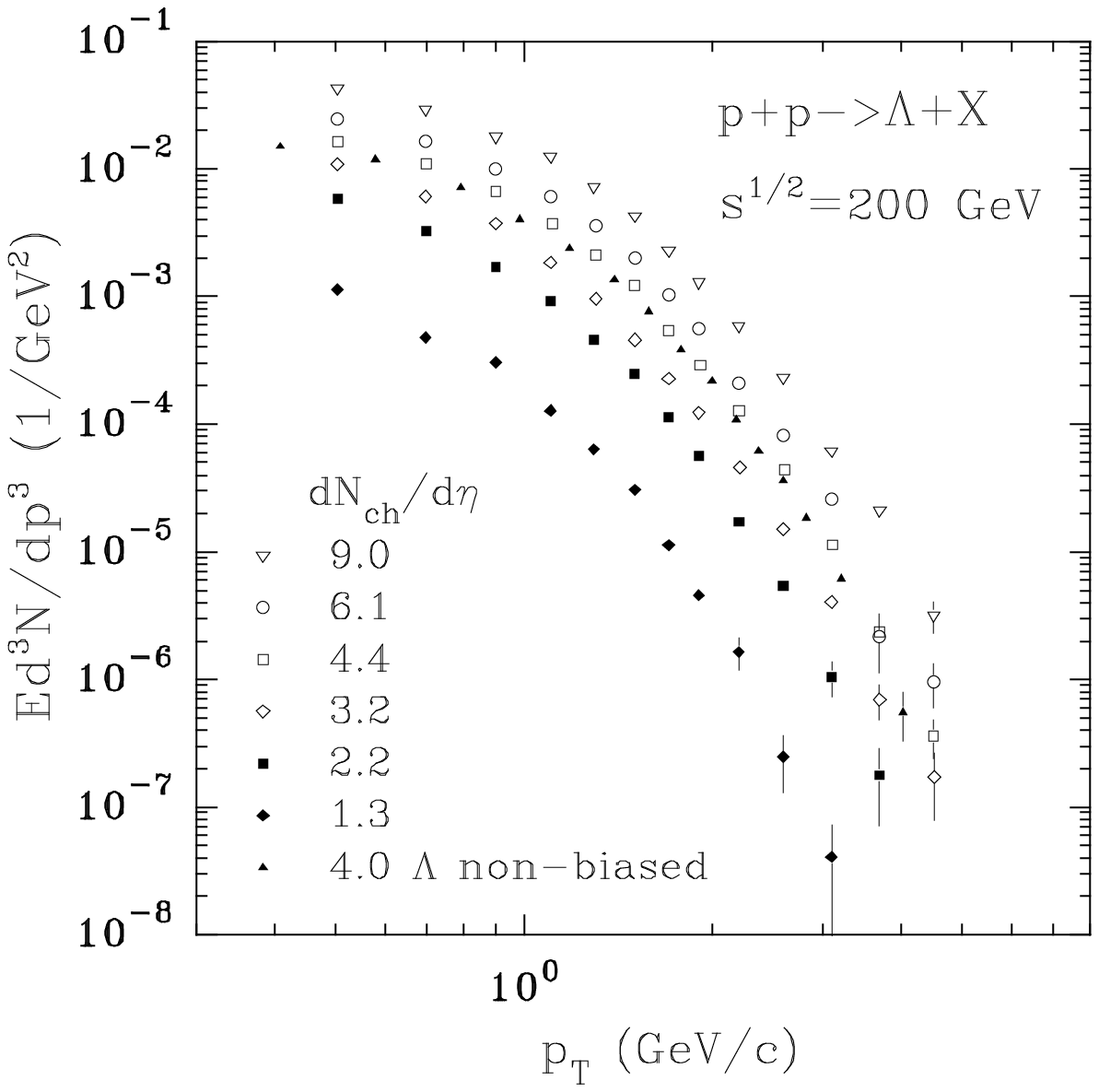}{}}
\hspace*{2.cm}
\parbox{6cm}{\epsfxsize=6.cm\epsfysize=6.cm\epsfbox[95 95 400 400]
{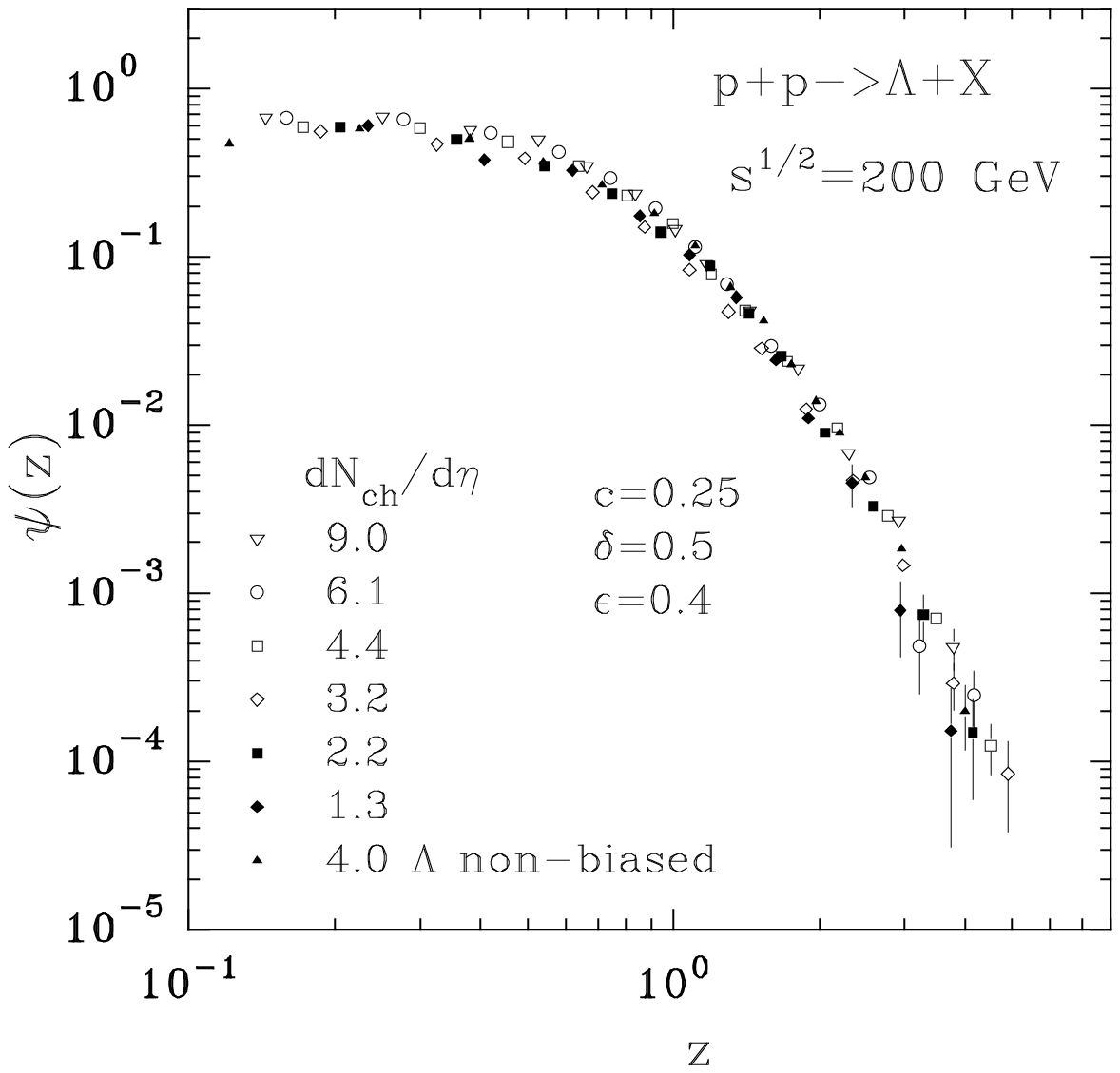}{}}
\vskip -2.cm
\hspace*{0.cm} a) \hspace*{8.cm} b)\\[0.5cm]
\end{center}

{\bf Figure 11.}
(a) Transverse momentum spectra of $\Lambda$-baryons produced
in $pp$ collisions for different multiplicity densities at $\sqrt s=200$~GeV.
The $\Lambda$-baryon spectrum for non-biased $pp$ collisions at $\sqrt s=200$~GeV
is shown by black triangles.
Experimental data are taken from Ref. \cite{Witt}.
(b) The corresponding scaling function $\psi(z)$.


\vskip 0.5cm
\begin{thebibliography}{99}


\bibitem{Quarkcom}
E. Eichten, K. Lane and M. Peskin, Phys. Rev. Lett. {\bf 50}, 811 (1983).\\
E. Eichten, I. Hinchliffe, K. Lane and C. Quigg, Rev. Mod. Phys.
{\bf 56}, 4 (1984).

\bibitem{Extradim}
I. Antoniadis, in {\it Proceedings of European School of
High-Energy Physics}, Beatenberg, Switzerland, 26 August - 8
September, 2001 (Editors: N.Ellis and J.March-Russul) p.301.

\bibitem{Blackhole}
C.G. Lester, in {\it Proceedings of Advanced Studies Institute on
"Physics at LHC"}, Czech Republic, Prague, July 6-12, 2003
(Editors: M Finger, A.Janata and M.Virius) A303.

\bibitem{Fracspace}
L. Nottale, {\it Fractal Space-Time and Microphysics} (World Sci., Singapore, 1993).\\
B. Mandelbrot, {\it The Fractal Geometry of Nature} (Freeman, San Francisco, 1982).

\bibitem{Feynman}
R.P. Feynman, Phys. Rev. Lett. {\bf 23}, 1415 (1969).

\bibitem{Bjorken}
J.D. Bjorken, Phys. Rev. {\bf 179}, 1547 (1969); J.D. Bjorken and
E.A. Paschanos, Phys. Rev. {\bf 185}, 1975 (1969).

\bibitem{Bosted}
P. Bosted {\it et al.}, Phys. Rev. Lett. {\bf 49}, 1380 (1972).

\bibitem{Benecke}
J. Benecke  {\it et al.},  Phys. Rev.  {\bf 188}, 2159 (1969).

\bibitem{Baldin}
A.M. Baldin,
Sov. J. Part. Nucl. {\bf 8}, 429 (1977).

\bibitem{Stavinsky}
 V.S. Stavinsky,
Sov. J. Part. Nucl. {\bf 10}, 949 (1979).


\bibitem{Leksin}
G.A. Leksin: Report No. ITEF-147, 1976; G.A. Leksin: in
{\it Proceedings of the XVIII International Conference on High
Energy Physics}, Tbilisi, Georgia, 1976, edited by N.N.
Bogolubov {\it et al.}, (JINR Report No. D1,2-10400, Tbilisi,
1977), p. A6-3.

\bibitem{KNO}
Z. Koba, H.B. Nielsen and P. Olesen,
Nucl. Phys.  {\bf B40}, 317 (1972).

\bibitem{Matveev}
  V.A. Matveev, R.M. Muradyan and A.N. Tavkhelidze,
  Part. Nuclei {\bf 2}, 7 (1971);
  Lett. Nuovo Cim. {\bf 5},  907 (1972);
  Lett. Nuovo Cim. {\bf 7}, 719 (1973).

\bibitem{Brodsky}
  S. Brodsky and G. Farrar,
  Phys. Rev. Lett. {\bf 31}, 1153 (1973);
  Phys. Rev.  {\bf D11}, 1309 (1975).


\bibitem{Z}
I. Zborovsk\'{y}, Yu.A. Panebratsev, M.V. Tokarev and G.P. \v{S}koro,
Phys. Rev. {\bf D 54}, 5548 (1996); I. Zborovsk\'{y}, M.V. Tokarev,
Yu.A. Panebratsev and G.P. \v{S}koro, Phys. Rev. {\bf C59}, 2227
(1999); M.V. Tokarev and T.G. Dedovich, Int. J. Mod. Phys. {\bf A15},
3495 (2000); M.V. Tokarev, O.V. Rogachevski and T.G. Dedovich, J.
Phys. G: Nucl. Part. Phys. {\bf 26}, 1671 (2000); M.V. Tokarev,
O.V. Rogachevski and T.G. Dedovich, Preprint No. E2-2000-90, JINR
(Dubna, 2000); M. Tokarev, I. Zborovsk\'{y}, Yu. Panebratsev and
G. Skoro, Int. J. Mod. Phys. {\bf A16}, 1281 (2001); M. Tokarev,
hep-ph/0111202; M. Tokarev and D. Toivonen, hep-ph/0209069;
G.P. Skoro, M.V. Tokarev, Yu.A. Panebratsev and I. Zborovsk\'{y},
hep-ph/0209071; M.V. Tokarev, G.L. Efimov and D.E. Toivonen, Physics
of Atomic Nuclei, {\bf 67}, 564 (2004); M. Tokarev, Acta Physica
Slovaca, {\bf 54}, 321 (2004); M.V. Tokarev and T.G. Dedovich,
Physics of Atomic Nuclei, {\bf 68}, 404 (2005).

\bibitem{ZZ}
I. Zborovsk\'{y} and M.V. Tokarev, hep-ph/0506003, to appear in
Part. and Nucl., Letters.

\bibitem{SRT}
I. Zborovsk\'{y}, hep-ph/0311306; I. Zborovsk\'{y}, in
{\it Proceedings of the XVII Inernational Baldin Seminar on High Energy Physics
Problems}, Dubna, Russia, 2004, edited by A.N. Sissakian, V.V. Burov and
A.I. Malakhov, Vol.I., p.167-172.


\bibitem{FNAL1}
D. Antreasyan {\it  et al.}, Phys. Rev. {\bf D19}, 764 (1979).

\bibitem{ISR}
BS Collaboration, B. Alper {\it  et al.}, Nucl. Phys. {\bf B100}, 237 (1975).

\bibitem{CDHW}
CDHW Collaboration, D. Drijard {\it  et al.}, Nucl. Phys. {\bf B208}, 1 (1982).

\bibitem{FNAL2}
D.E. Jaffe {\it  et al.}, Phys. Rev. {\bf D40}, 2777 (1989).

\bibitem{STAR1}
STAR Collaboration, J. Adams {\it  et al.}, Phys. Rev. Lett. {\bf 91}, 172302 (2003).

\bibitem{Baranikova}
O. Baranikova (STAR Collaboration), in {\it Proceedings of the
Quark Matter 2005}, Aug. 4-5, 2005, Budapest, Hungary; J. Adams
{\it  et al.}, nucl-ex/0601033.

\bibitem{STAR2}
STAR Collaboration, J. Adams {\it  et al.}, Phys. Lett. {\bf B616}, 8 (2005).

\bibitem{Adams}
J. Adams and M. Heinz for the STAR Collaboration, nucl-ex/0403020.


\bibitem{Ward}
D.R. Ward, Report No. CERN-EP/87-178, 1978 (unpublished);
W. Thom\'{e} {\it  et al.}, Nucl. Phys. {\bf B129}, 365 (1977).

\bibitem{CHLM}
CHLM Collaboration, M.G. Albrow {\it  et al.},  Nucl. Phys. {\bf B56},333-345 (1973).

\bibitem{Gans}
J.E. Gans, PhD Thesis, Yale University, USA, 2004.

\bibitem{Witt}
R. Witt for the STAR Collaboration, J. Phys. G: Nucl. Part. Phys. {\bf 31}, S863-S871 (2005).






\end{thebibliography}
\end{document}